# A Secured Energy Saving with Federated Assisted Modified Actor-Critic Framework for 6G Networks

Attai Ibrahim Abubakar*†, Michael S. Mollel†‡, Metin Ozturk§, Naeem Ramzan*

*Abstract*—**Ultra-dense base station deployments must be operated in a way that allows the network's energy consumption to be adapted to the spatio-temporal traffic dynamics, thereby minimizing overall energy consumption. To achieve this goal, we leverage two artificial intelligence algorithms—federated learning and actor-critic—to develop a proactive and intelligent cell switching framework. This framework can learn the operating policy of small base stations in an ultra-dense heterogeneous network, resulting in maximum energy savings while respecting quality of service (QoS) constraints. Additionally, the use of federated learning enhances the security of the system as only the model parameters are shared among the entities, rather than the raw and potentially sensitive data. In other words, in the event of a data breach or eavesdropping, only the parameters of the developed artificial intelligence model would be compromised. This approach ensures that the network remains secure against attacks targeting confidential and sensitive data. The performance evaluation reveals that the proposed framework can achieve an energy saving that is about 77% more than that of the state-of-the-art solutions while respecting the QoS constraints of the network.**

## I. Introduction

Cell switching, which involves turning base stations (BSs) on or off dynamically, is one of the most effective solutions for maximizing energy savings in cellular networks. This approach aims to adapt the energy consumption of BSs to the varying traffic demand and can be implemented with minimal modifications to the existing network architecture [1], [2]. However, the challenges involved in implementing this approach from an algorithmic perspective have been a major hindrance preventing its adoption in real-life networks. It has been proven in [3] that the BS switching problem is a combinatorial optimization problem, which is NP-hard. This is because the number of possible search spaces to select the optimal solutions grows exponentially with the number of BSs deployed in the network. As such, for large-scale network deployment, the number of search spaces can grow extremely large, making it difficult to find suitable algorithms that can be used to find the optimal solutions for such combinatorial problems.

Exhaustive search (ES), also known as brute force, is the only algorithm that is always guaranteed to find the optimal solution. However, it encounters substantial computational overhead when the network size becomes enormous, and as such, it is only suitable for small-size network deployments [4]. Hence, a near-optimal solution with much less computational overhead is often resorted to, enabling the solution to scale with the network size. In this regard, heuristics, meta-heuristic, and machine learning-based solutions have been proposed to find near-optimal solutions [3], [5]–[23]. However, some of the solutions developed still suffer from colossal computation overhead and are not adaptable to dynamically changing network traffic conditions. In addition, most of them require regular sharing of traffic information among the BSs to find the optimal solution, which raises security concerns and the possibility of compromising the data privacy of the users in the network. There is also the problem of high signaling overhead which leads to increased bandwidth consumption [24], [25].

To tackle these challenges, we propose a reinforcement learning (RL)-based energy optimization framework, called federated-assisted modified actor-critic (FAMAC), for ultra-dense 6G networks. This framework involves two stages: In the first stage, federated learning (FL) is applied to determine the traffic prediction model of each small BS (SBS) in the network. These prediction models are then shared with a central controller, which constructs a look-up table with the models where the instantaneous traffic demands of each SBS can be easily obtained, rather than having each SBS continually send its traffic demand to the macro BS (MBS) whenever a switching decision is to be made. The second stage involves learning the optimal switching strategy to minimize energy consumption using a modified version of actor-critic deep RL (DRL). In this approach, we adjust the internal architecture of the actor to enable it to handle problems with large action spaces, such as cell switch off/on in ultra-dense networks.

Using FL in the first stage enhances network security as FL requires sharing only model parameters instead of raw user traffic data, thereby alleviating privacy concerns. Different levels of privacy issues arise depending on the type of data shared and the design of the algorithm used. If, for instance, the algorithm requires only high-level traffic information of SBSs, privacy concerns exist at the network level since the network's sensitive data can be compromised. However, if the shared data includes more contextual information about users in an anonymous manner, the privacy issues persist still at the network level, as the type of traffic at specific SBSs is revealed even though the users remain unidentified. Lastly, if user content is shared without anonymization, it poses critical privacy issues from both network and user perspectives, potentially leading to serious consequences by breaching confidential and sensitive data. In addition, we introduce both global and local weights, where the former facilitates cooperation among the various SBSs' traffic parameters, while the latter learns features specific to the network conditions at individual SBSs. A key novelty in this work is a redesigned actor network that can directly output optimal BS switch off decisions for individual SBSs in ultra-dense heterogeneous networks (UDHNs). This effectively transforms the conventional exponential joint action space into a simplified linear output, overcoming the scalability issues of traditional actor-critic methods. By enabling efficient exploration of massive discrete joint actions, the customized actor architecture allows the learning of optimized cell switching policies across the vast combinatorial search space. This significant contribution enables the scalable application of actor-critic algorithms for data-driven and intelligent energy optimization in next-generation communication systems.

*School of Computing, Engineering and Physical Sciences, University of the West of Scotland, Paisley, UK. †James Watt School of Engineering, University of Glasgow, Glasgow, UK. ‡Nelson Mandela African Institute of Science and Technology Arusha, Arusha, TZ. §Electrical and Electronics Engineering, Ankara Yıldırım Beyazıt University, Ankara, Türkiye. Corresponding email: Naeem.Ramzan@uws.ac.uk.

TABLE I
TABLE OF NOTATIONS

| Symbol | Description |
|---|---|
| $\Phi$ | Data rate |
| $G$ | Number of resource blocks |
| $\kappa_{\text{RB}}$ | Bandwidth of one resource block |
| $\lambda$ | Signal-to-noise-plus-interference ratio (SINR) |
| $\psi$ | BS load |
| $\Lambda$ | Number of users |
| $H_j$ | Total number of resource blocks |
| $P_{\text{o}}$ | Constant power consumption |
| $P_{\text{s}}$ | Sleep mode power consumption |
| $\mu$ | Power amplifier efficiency |
| $P_{\text{tx}}$ | Transmit power |
| $Z_{\text{T}}$ | Total number of time slots |
| $Z_{\text{m}}$ | Total number of macro cells |
| $Z_{\text{s}}$ | Total number of SBSs |
| $Z_{\text{b}}$ | Total number of BSs |
| $\Gamma_t$ | OFF/ON status of BS |
| $\Theta^z$ | Global actor network weights |
| $\theta_n$ | Local actor network weights |
| $\Theta^p$ | Combined local and global weight |
| $\alpha^\theta$ | Local actor's learning rate |
| $\zeta$ | Learning rate for the global network |

### A. Contributions

This work proposes a DRL framework, referred to as FAMAC, for secure energy saving in ultra-dense 6G networks. The main contributions of this work are as follows:

- We employ FL to develop a traffic prediction model based on a real network dataset to preserve the privacy of user traffic while minimizing the signaling overhead incurred due to information exchange between SBSs and the MBS.
- We present the mathematical model for computing the power consumption of the ultra-dense network and formulate the SBS switching problem as a constrained combinatorial optimization problem.
- We develop a modified actor-critic framework to determine the optimal switch off/on pattern of the SBSs in the network while respecting the quality of service (QoS) constraints. We also build a set of procedures that would enable the proposed framework to overcome the limitations of conventional actor-critic algorithms in handling combinatorial problems with large discrete action space and multiple constraints.

## II. RELATED WORKS

Several algorithmic approaches have been developed by researchers to solve the complex problem of finding the optimal or near optimal cell switching solution to minimizing the energy consumption of cellular networks. These techniques can be broadly grouped into two categories: non-machine learning and machine learning techniques. The non-machine learning approaches include heuristics, such as greedy algorithm, sorting algorithms, and other local search methods and meta-heuristics, such as genetic algorithm, simulated annealing, particle swarm optimization, etc. The machine learning techniques include reinforcement and deep learning that have been proven to learn from data and adapt to a dynamically changing network environment.

### A. Non-Machine Learning Approaches

For the non-machine learning approaches, the work in [26] introduced a dynamic and distributed cell switching scheme where the criteria for switching off SBSs is based on a set of key parameters including user density variation and random spatial network model. The authors in [27] proposed a threshold-based cell switching algorithm in cellular networks. This approach employs a threshold parameter to determine the maximum transmission power limit for active BSs. By establishing this threshold, the algorithm ensures that the QoS for users remains uncompromised during cell switch-off operations, effectively balancing energy efficiency with network performance. In [28], a meta-heuristic algorithm that exploits landscape characteristics to improve a multi-objective function was developed to solve the cell switching problem in 5G ultra-dense networks. The application of simulated annealing algorithms for cell switching in heterogeneous cellular networks has been investigated in [5]–[7]. Additionally, particle swarm optimization techniques have been employed to derive near-optimal cell switching solutions [8]–[10]. Other heuristic approaches, such as sorting-based algorithms, have been explored in [11], [12], wherein the traffic load of SBSs is ranked in ascending or descending order, with those exhibiting the lowest traffic load selected for deactivation. Furthermore, genetic algorithms have been utilized to address the cell switching problem in multiple investigations [13], [14], [29].

The problems with heuristic and meta-heuristic approaches are that they (i) are executed with conditional statements, (ii) are difficult to generalize to unseen observations, and (iii) cannot easily adapt to a dynamically changing network condition as obtained in ultra-dense 6G networks. In addition, due to the dynamic changes in wireless communication networks, there is a need to apply these solutions to calculate the optimal solution each time the network condition changes, resulting in increased computational overhead [30], [31].

### B. Machine Learning Approaches

The other approach to cell switching is machine learning solutions. Among the machine learning techniques, several works on cell switch off/on have been implemented using RL in the literature. For example, in [15]–[17], $Q$-learning frameworks for cell switching and traffic offloading in a heterogeneous network (HetNet) were proposed. In these works, the HetNets comprise an MBS and several SBSs, and the learning takes place at the MBS, which serves as the controller of all the BSs under its coverage. The drawback of $Q$-learning algorithm is that it involves the use of tables ($Q$-tables) to store the learnt state-action values ($Q$-values), thereby in situations where a large number of actions need to be learnt, it would take very long for the $Q$-table to converge. In addition, an extensive memory would be required to store the learnt $Q$-table, and as a result, with increasing actions or states, the generalization ability of $Q$-learning begins to diminish [32]. Hence, the tabular approaches of RL are only feasible when the network dimension is small.

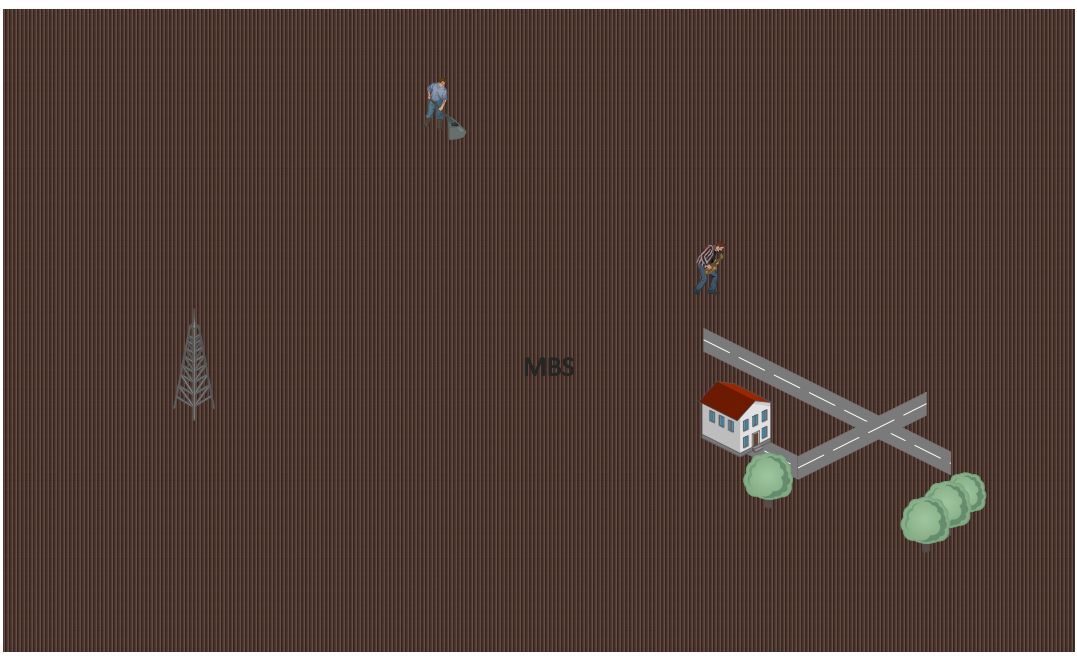

Fig. 1. The network model consists of an MBS and various types of SBSs.

To address the curse of dimensionality inherent in conventional and tabular learning techniques, the use of function approximators has been proposed [32]. In this context, the authors in [3] extended the work of [15] and [16] by introducing SARSA algorithm with value function approximation. This approach enhances the scalability and generalization capabilities of the conventional SARSA algorithm, enabling its application to large-scale networks. Comparable efforts to improve the scalability of $Q$-learning models have been undertaken in [18], [19], [33], [34]. These studies employed deep $Q$-networks and double deep $Q$-networks to address scalability issues. While deep $Q$-networks generally demonstrate the ability to handle continuous states and action spaces, their performance has been observed to deteriorate when confronted with extensive discrete action spaces [32].

Actor-critic [20], [21] and deep-actor-critic [22], [23] RL models have also been employed for cell switching. The conventional actor-critic algorithm is unsuitable for application in problems involving large state or action spaces because the critic employs a state-action value table as its value function, requiring ample storage for large-scale problems [22]. Similarly, in the conventional deep actor-critic algorithm, the output of the actor-network is a probability distribution of possible actions and is only feasible when a few numbers of action spaces are considered. Hence, the actor-critic and deep actor-critic solutions are only suitable for small- to medium-scale network deployments, and there is still the need for a scalable solution that is adaptable to any network size to be developed.

Therefore, in this work, we develop a solution that is able to overcome the challenges of handling large state spaces as well as huge discrete action spaces facing both conventional actor-critic and actor-critic DRL models. Specifically, we leverage FL alongside the actor-critic RL algorithm to develop a secure energy-saving framework for ultra-dense 6G cellular networks, referred to as FAMAC, which is able to learn the optimal cell switch off/on policy even when the number of SBSs in the network becomes very large while incurring minimal signaling overhead compared to other state-of-the-art solutions. The proposed model is trained in an offline manner, after which the trained model is implemented in real-time for energy saving in the network.

## III. System Model

### A. Network Model

As illustrated in Fig. 1, a UHDN consisting of several macro cells (MCs) is considered where each MC comprises an MBS, and several SBSs (remote radio head (RRH), micro, pico and femto) having varying capacities and power consumption profiles. The architecture considered for the UDHN is the separated control and data architecture where the MBSs function as control BSs for the purpose of signalling, low data rate transmission, and activation and deactivation of the SBSs deployed in their coverage. The SBSs function as data BSs and are deployed in hotspot areas for the purpose of capacity and throughput enhancement. In addition, vertical traffic offloading is employed whereby the traffic load of switched-off SBSs can be transferred to the MBS. The goal of traffic offloading is to ensure that the traffic demand of the users connected initially to the switched-off SBSs are served by the MBS to maintain the QoS of the network. It is assumed in this work that prior to the cell switch off/on operation, the UDHN has sufficient radio resources to serve all user demands when all the BSs in the network are active. However, this cannot be guaranteed when the cell switch off/on scheme is implemented since the number of radio resources in the network will reduce when some SBSs are switched off to save energy. In addition, the MCs are assumed to function in a decentralized manner such that the MBS in each MC is saddled with the responsibility of controlling the activities of the SBSs in its coverage.

### B. Base Station Load

The radio resources of a BS are segmented into units, i.e., resource blocks (RBs), and to meet the QoS requirements, a certain amount of RBs is assigned to each user. Hence, it is necessary to determine the minimum number of RBs that is required to meet the data rate[1] ($\Phi_k$) of the user $k$ in order to guarantee that its QoS is satisfied by the network. As a result, the number of RBs, $G_{k,j}$, that must be assigned to user $k$ by BS $j$ is given by $G_{k,j} = \frac{\Phi_k}{\kappa_{\text{RB}} \log_2(1+\lambda_{k,j})}$, where $\kappa_{\text{RB}}$ is the bandwidth of one RB and $\lambda_{k,j}$ is the signal-to-noise-plus-interference-ratio (SINR) received by user $k$ from BS $j$. The BS load can then be defined as $\psi_j = \frac{\sum_{k=1}^{\Lambda} G_{k,j}}{H_k}$, where $\Lambda$ is the number of users connected to BS $j$ and $H_j$ is the total number of RBs in BS $j$.

### C. Power Consumption of UDHN

The total power consumption of the UDHN comprises the power consumption of the several MCs in the network, each containing an MBS and several SBSs. We represent BS $j$ in MC $i$ by $\beta^{i,j}$ where $\beta^{i,1}$ is used to identify the MBS. According to the EARTH model, the instantaneous power consumption of a BS, $P_{\text{BS}}$, can be expressed as [35], [36]

$$P_{\text{BS}} = \begin{cases} P_{\text{o}} + \mu P_{\text{tx}}, & 0 < P_{\text{tx}} < 1 \\ P_{\text{s}}, & P_{\text{tx}} = 0, \end{cases} \tag{1}$$

where $P_{\text{o}}$ and $P_{\text{s}}$ are the constant and sleep power consumption component, respectively, $\mu$ denotes the power amplifier efficiency, and $P_{\text{tx}}$ is the transmit power, such that

$$P_{\text{tx}} = \begin{cases} \psi P_{\text{max}}, & 0 < \psi < 1 \\ P_{\text{max}}, & \psi = 1, \end{cases} \tag{2}$$

where $P_{\text{max}}$ is the maximum allowable transmit power. We consider all SBSs of the same category to have similar hardware, such that $\mu$ and $P_{\text{o}}$ are the same. We do not consider power allocation thereby,

[1]The term *data rate* here represents the maximum data rate (i.e., channel capacity) over a noisy wireless communication channel.

the transmit power of each type of BS is fixed among BS belonging to the same category.

Combining (1) and (2), the instantaneous power consumption of BS $j$ in MC $i$ at time $t$, $P^{i,j}_{\beta,t}$, can be expressed as $P^{i,j}_{\beta,t}(\psi_t) = P^{i,j}_{\text{o}} + \psi^{i,j}_t \mu^{i,j} P^{i,j}_{\text{max}}$. Therefore, the total instantaneous power consumption of an ultra-dense 6G cellular network, $P_{\text{total}}$, with several MCs and different types of BSs at $t$ can be expressed as $P_{\text{total}}\ (\psi^{i,j}_t) = \sum_{i=1}^{Z_{\text{m}}} \sum_{j=1}^{Z_{\text{b}}} P^{i,j}_{\beta,t}(\psi^{i,j}_t)$, where $P^{i,1}_t$ represents the power consumption of MBS in $i^{\text{th}}$ MC, $Z_{\text{m}}$ is total number of MCs in the network, and $Z_{\text{b}}$ is the total number of BSs in an MC, where $Z_{\text{b}} = Z_{\text{s}} + 1$ with $Z_{\text{s}}$ being the number of SBSs within an MC.

## IV. Problem Formulation and Optimization Objective

### A. Problem Formulation

A certain interval of time ($T$) is considered such that $T$ is subdivided into various time slots (in mins) which are of the same duration $L$ (in mins). We also define an index vector $\nu$, storing all the time slots in successive order, that is given as $\nu = [1,2,...,Z_{\text{T}}]$, where $Z_{\text{T}}$ denotes the number of time slots and is given by $Z_{\text{T}} = T/L$. The scenario considered in this work is one where the UDHN can decide to switch off/on a certain number of SBSs when the traffic load of the network is low to reduce the amount of energy consumption in the network. We aim to determine the optimal switching policy (i.e., the optimal combination of SBSs to switch off/on) at each time slot that maximizes energy savings.

Hence, the total power consumption of the UDHN (accumulated over all the time slots) that would be obtained with the implementation of cell switch off/on can be expressed as

$$P_{\text{total}}\ (\psi^{i,j}_t, \Gamma^{i,j}_t) = \sum_{\nu=1}^{Z_{\text{T}}} \sum_{i=1}^{Z_{\text{m}}} \sum_{j=1}^{Z_{\text{b}}} [\Gamma^{i,j}_t P^{i,j}_{\beta,t}(\psi^{i,j}_t) + (1-\Gamma^{i,j}_t) P^{i,j}_s], \tag{3}$$

where $\Gamma^{i,j}_t$ denotes the off/on status of BS $j$ in MC $i$ at $t$, i.e.,

$$\Gamma^{i,j}_t = \begin{cases} 1, & \text{if } \beta^{i,j} \text{ is ON} \\ 0, & \text{if } \beta^{i,j} \text{ is OFF.} \end{cases} \tag{4}$$

Note that since the MBS is always active, $\Gamma^{i,1}_t = 1, \forall t$.

### B. Optimization Problem Modeling

The optimization objective is to minimize the total power consumption of the UDHN while satisfying the QoS requirement of the network. The main source of power consumption that needs to be minimized is BSs. However, cell switching strategies may result in additional power consumption due to transitions between active and sleep states. When the problem is considered holistically, several variables must be accounted for in calculating the total power consumption. The notable components of total power consumption are: (i) the power consumption of BSs during their active and sleep states, (ii) the transient power consumption of BSs during their state changes, and (iii) the power consumption of UEs during handovers when their serving BSs are switched off. Here, the minimization optimization objective can be defined as

$$\min \quad P_{\text{total}}(\psi^{i,j}_t, \Gamma^{i,j}_t) + P_{\text{tran}} + P_{\text{ho}}, \tag{5}$$

where $P_{\text{tran}}$ and $P_{\text{ho}}$ are the transition and handover power consumption, respectively.

It is worth noting that frequent handovers not only increase user-side energy consumption but also degrade the QoS of users [37]. However, since we are interested in network-side power consumption, we do not take into account the handover power consumption, $P_{\text{ho}}$. Moreover, we acknowledge that transition power consumption, $P_{\text{tran}}$, is an important aspect to investigate, as it can reduce total energy savings when the frequency of transitions for a certain BS increases. However, since it requires dedicated attention and separate modeling, it is beyond the scope of this current work. Therefore, the optimization problem in (5) is updated as

$$\min_{\Gamma^{i,j}_t} \quad P_{\text{total}}(\psi^{i,j}_t, \Gamma^{i,j}_t) \tag{6}$$

$$\text{s.t.} \quad \Phi_i = \chi_i, \quad \forall i,j, \tag{7}$$

$$\hat{\psi}^{i,1}_t \leq \psi^{i,1}_{t,\text{m}}, \quad \forall i, \tag{8}$$

$$\Gamma^{i,j}_t \in \{0,1\}, \quad \forall i,j. \tag{9}$$

The total traffic demand of the MC, $\Phi_i$, when all the BSs in the MC $i$ are switched on (i.e., prior to traffic offloading) is expressed as $\Phi_i = \sum_{j=1}^{Z_{\text{b}}} \psi^{i,j}_t$, $\forall i$. To satisfy the QoS requirement, we try to ensure that the traffic demand of any SBS that is switched off will be offloaded to the MBS, hence, the actual traffic load of MBS during the process of traffic offloading which is denoted by $\hat{\psi}^{i,1}_t$ can be expressed as $\hat{\psi}^{i,1}_t = \psi^{i,1}_t + \sum_{j=2}^{Z_{\text{b}}} \psi^{i,j}_t (1-\Gamma^{i,j}_t), \quad \forall i$. When traffic offloading is completed, the traffic demand of the MC, $\chi_i$, can be expressed as $\chi_i = \hat{\psi}^{i,1}_t + \sum_{j=2}^{Z_{\text{b}}} \psi^{i,j}_t \Gamma^{i,j}_t$, $\forall i$.

There is a maximum limit to the amount of traffic demand that can be accommodated by the MBS (i.e., $\psi^{i,1}_{t,\text{m}}$), due to the amount of radio resources available at the MBS. Therefore, an additional constraint is defined in (8).

## V. Proposed Methodology

The implementation procedure for the proposed FAMAC framework for cell switching in 6G networks comprises two stages. The first stage is the traffic prediction stage, where an FL model is developed to equip the MBS with a global knowledge of the traffic load of all the SBSs under its coverage. The second stage is the cell switching stage, where a modified actor-critic DRL model is developed to determine the optimal cell switching policy of the UDHN. Random state changes are typically involved in the second stage (especially when RL is employed) due to the random search mechanisms, i.e., exploration, during algorithm training. It is hard, if not impossible, to implement these random state changes in real networks. However, with the help of the first stage, this becomes possible. The traffic loads of BSs, which are the main determining factor in the second stage, are predicted, subsequently facilitating the second stage by enabling early implementation and preparation.

### A. Traffic Prediction Stage

In the traffic prediction stage, we leverage the concept of federated machine learning to obtain the future traffic loads of all the SBSs in the network. The primary motivation for employing FL in this work is to equip the decision-making entity, located at the MBS where the proposed FAMAC framework is implemented, with global knowledge of the traffic conditions of all the SBSs under its coverage. This comprehensive understanding of network-wide traffic patterns is crucial for making informed and optimized cell switching decisions.

In the FAMAC framework, the cell switching off/on decisions are made by a central server situated at the MBS. Without the use of FL for traffic prediction, each SBS would be required to periodically send its traffic condition information to this central server. This approach would result in several drawbacks, including increased signaling overhead, higher latency, and additional energy consumption associated with the frequent data transmissions. Furthermore, the direct sharing of traffic data from the SBSs to the MBS raises privacy concerns, as sensitive user information could potentially be compromised by unauthorized third parties during the data transfer process. To address these challenges, we harness the power of FL in the FAMAC framework. Instead of directly transmitting traffic data, each SBS trains a local traffic prediction model using its own historical data. The SBSs then share only the encrypted versions of the trained local model weights with the central server at the MBS. By aggregating these local model weights, the MBS gains a global understanding of the traffic conditions across all SBSs without requiring access to the raw traffic data. This approach significantly reduces the signaling overhead and latency associated with data transmission while ensuring the privacy of sensitive user information.

*1) FL-based Traffic Prediction Model:* The FL is a machine-learning technique that enables models to be trained collaboratively across multiple devices or network entities in different locations without sharing their data. In other words, in FL, the data used for training the model on individual devices is retained locally, and only the updates to the parameters of the local models are shared with the central learning entity, which aggregates all the updates to obtain a global model. The global model is then shared with the individual devices, and this process is repeated until convergence or the expected performance is attained. The procedure for implementing FL is discussed in the following paragraphs [38], [39].

We consider an MBS equipped with a server and $Z_s$ SBSs serving as its clients. Each SBS $n$ would utilize its training dataset $\varphi_n$ that is defined as $\varphi_n = \{x_{n,v}, y_{n,v}\}_{v=1}^{V}$, where $x_{n,v}$ denotes the neural network input of client $n$ and $y_{n,v}$ is the corresponding target. The main objective of the FL process is to minimize a loss function, which is defined as [40]

$$\min_{\tau \in \mathbb{R}^d} F(\tau) \equiv \frac{1}{\varphi} \sum_{n=1}^{Z_s} \varphi_n F_n(\tau), \tag{10}$$

where $F_n(\tau)$ is the loss function of each client $n$ (local loss function), $\tau \in \mathbb{R}^d$ is a vector of the parameters of the local model with dimension $d$, and $\varphi = \sum_{n=1}^{Z_s} \varphi_n$ is the sum of all the client's data samples. The step-by-step procedure given in [41] is followed, such that a back-and-forth communication with parameter updates is performed with the central entity (MBS), which stores a global model, and local clients (SBSs), storing local models, until the global model converges. This global model is then used to predict the future traffic load of the SBSs in the network. Since the prediction of future SBS traffic load is a time series problem, the FL model developed in this work utilized bi-directional long short-term memory (LSTM) due to its effectiveness in addressing such issues.

The bi-directional LSTM is an enhanced version of traditional LSTM, which was introduced to handle the problem of vanishing and exploding gradient that affects the performance of conventional recurrent neural networks (RNNs), thereby preventing them from learning long-term dependencies between data samples. The bi-directional LSTM improves the performance of the traditional LSTM by capturing the dependencies between past and future time steps, making them more suitable for sequence predictions such as time series forecasting, speech recognition, etc. It combines two traditional LSTM models: one processes information in the forward direction while the other does so in the backward direction, and the final output is obtained by concatenating the outputs from both LSTM models [42]. The pseudocode for the proposed FL-based traffic prediction is given in Algorithm 1.

**Algorithm 1** FL-Based Traffic Prediction Scheme

1: **Require:** Local Dataset $\varphi_n$, Number of SBS $Z_s$,
Number of federated rounds $Z_{\mathrm{f}}$,
Number of local Iteration $e$, Global Iteration T,
Learning rate $\eta$, local model parameter $\tau_n$
2: Initialize global weights $\Omega^0$
3: **for** $t=1$ to $T$ **do**
4: $j = \sum_{n \in Z_s} |\varphi_n|$
5: **for** $n \in Z_s$ **in parallel do**
6: Send the global model $\Omega^t$ to SBS $n$
7: $\Delta\tau_n^t \leftarrow$ LOCALTRAINING(n, $\Omega^t$)
8: **end for**
9: $\Delta_\Omega W(\Omega) \leftarrow \sum_{n \in Z_s} \frac{|\varphi_n|}{j} \Delta\tau_n^t$
10: $\Omega^{t+1} \leftarrow \Omega^t - \eta \Delta_\Omega W(\Omega)$ ($W(\Omega)$: is the cost function, $\Delta_\Omega W(\Omega)$: is the gradient of the cost function with respect to the parameters $\Omega$ )
11: **end for**
12: Global model $\Omega^T$ for traffic prediction.
13: **function** LOCALTRAINING($n$,$\Omega^t$)
14: $\tau_n^t = \Omega^t$
15: **for** $p=1$ to $e$ **do**
16: **for** each batch $\boldsymbol{b} = \{\boldsymbol{x}, y\}$ of $\varphi_n$ **do**
17: $\tau_n^t \leftarrow \tau_n^t - \eta \nabla L(\tau_n^t; \mathbf{b})$ ($\nabla L(\tau_n^t; \mathbf{b})$ is the gradient of the cost function with parameter $\tau_n^t$. Batch **b** contains independent variable **x** and dependent variable **y** )
18: $s.t \;\; L(\tau; \mathbf{b}) = \sum_{(x,y) \in \mathbf{b}} l(\tau; x; y)$
19: **end for**
20: **end for**
21: **return** $\tau_n^t$ to the MBS
22: **end function**

## B. Cell Switching Stage

In the cell switching stage, the future traffic loads of the network that were obtained from the traffic prediction stage are fed into the modified actor-critic model to determine the optimal policy in terms of the set of SBSs to be switched off to minimize the energy consumption of the network.

*1) Modified Actor-Critic Deep Reinforcement Learning:* Actor-critic is a class of RL algorithms where the agent comprises two parts, an actor and a critic. On one hand, the actor takes actions by following a policy $\pi(a_t|s_t)$, where $\pi(a_t|s_t)$ is the conditional probability density function. The critic, on the other hand, evaluates the decisions taken by the actor using a $Q$-value which can be expressed as $Q^\pi(s_t, a_t) = \mathbb{E}_{a_t \sim \pi(a_t)|s_t)} [R_t \,|\, s_t, a_t]$, where $\mathbb{E}[R_t \,|\, s_t, a_t]$ denotes a conditional expectation with respect to the policy $\pi(s_t \,|\, a_t)$, $R_t$ represents the cumulative discounted reward.

In contrast to conventional actor-critic algorithms, the proposed modified actor-critic DRL framework employs a centralized learning system where decisions regarding the switching policy of all SBSs in the network are managed by a central entity, specifically the MBS. This central entity possesses comprehensive information about the network's traffic conditions across all SBSs, enabling it to derive optimal switching policies based on the FL model.

The major modification is with respect to the actor network. Usually, the output of the actor-network of a conventional actor-critic DRL algorithm is a probability distribution comprising all possible actions that can be taken. However, in our work, since the number of actions scales exponentially with the number of SBSs ($2^{(Z_s-1)}$), the algorithm would require a very large number of neurons at the output. In addition, the huge action space may lead to inefficient exploration of the action space during the learning process, thereby degrading the overall performance of the algorithm.

To overcome the aforementioned challenges, we redesign the actor network in such a way that it can output a single deterministic action vector to represent the operating condition of each SBS in the network. To achieve this, we replace the softmax function, which is responsible for generating the exponential action space, with a sigmoid function. Thus, we are now able to reduce the action space from exponential to linear, which improves the efficiency of exploration and dramatically enhances the overall performance of the proposed method. Therefore, in our modified approach, the actor learns the individual switching policies of the SBSs separately using the sigmoid function, after which the outputs from each sigmoid function are combined as the actor's output.

The actor network is divided into two parts: global weights ($\Theta^z$), which are shared among all SBSs, and local weights ($\theta_n$), which are specific to each SBS. Using sigmoid functions for individual switching decisions reduces the action space from exponential to linear, enhancing efficient exploration and scalability. The global weights, updated based on aggregated gradients, promote cooperation and shared learning among SBSs, enabling coordinated decision-making. This allows the joint action space to be explored efficiently, as the switching-off action of one SBS affects the options of others. The local weights are updated separately using only the local gradient for each SBS, allowing for specialized adaptations to each SBS's traffic patterns and network environment. This combination of shared and private parameters balances coordinated decision-making with personalized policy learning, improving overall performance. The learning process is further supported by the critic network, which provides comprehensive feedback. In essence, the global weights foster cooperation between SBS switching off policies, while the local weights capture specialized behaviour, enabling the actor-network to handle the large joint action space and learn coordinated yet personalized policies.

Although it may appear that multiple entities are collaborating to minimize the overall power consumption of the UDHN, the proposed approach remains centralized. The central entity, the MBS, is solely responsible for managing the decision-making process as the only agent within the proposed DRL scheme. However, the overall FAMAC framework involves multiple entities, including the MBS and SBSs, working together to facilitate the cell switching process, which could cause some confusion regarding the algorithmic architecture. To clarify and avoid possible confusion, while the FAMAC framework consists of multiple entities involved in data provision, traffic prediction, and decision-making, the DRL scheme is exclusively managed by the centralized entity—the MBS.

The loss function of the conventional actor-critic DRL, $-J(\theta_t)=-\mathbb{E}\left[Q^{\theta}(s_t,a_t;\omega_t)\right]$,—$\omega_t$ represents the learned parameters of the function—is modified to reflect the combination of sigmoid outputs for each SBS in the proposed solution. Modification is performed by replacing the weight $\theta \rightarrow [\Theta^z;\ \theta_n] \rightarrow \Theta^p$, where $p$ represents the combined weight of the global and the local network and $\Theta^z$ is global weight connected to the $n$-th SBS. Therefore, the loss gradient of actor becomes $\nabla_{\Theta} J(\Theta^p)$, which can be written as

$$\begin{aligned}\nabla_{\boldsymbol{\Theta^p}} J(\boldsymbol{\Theta^p_t}) &= \mathbb{E}[\nabla_{\boldsymbol{\Theta^p}} \log(\pi(\boldsymbol{a}_t \,|\, \boldsymbol{s}_t;\boldsymbol{\Theta^p_t}))\delta_V(\boldsymbol{\omega}_t)], \\ &\approx \frac{1}{T}\sum_{t=1}^{T}[\nabla_{\boldsymbol{\Theta^p}} \log(\pi(\boldsymbol{a}_t \,|\, \boldsymbol{s}_t;\boldsymbol{\Theta^p_t}))\delta_V(\boldsymbol{\omega}_t)],\end{aligned} \tag{11}$$

where $\nabla_{\boldsymbol{\Theta^p}} J(\boldsymbol{\Theta^p_t})$ represents the gradient of the actor loss for an individual network of $n \in Z_{\text{s}}$ SBSs which contains global weight $\Theta^z$ and local weights $\theta_n$, $M$ is the number of collected transitions. In this work, we use TD(0), so $\pi(\boldsymbol{a}_t \,|\, \boldsymbol{s}_t;\boldsymbol{\Theta^p_t})$ is the predicted probability of selecting action $a_t$ given state $s$ for the $m$-th transition in the network and $\delta_V(\boldsymbol{\omega}_t)$ is the advantage value or TD for the $m$-th transition.

The update rule of $\Theta^p$ can be derived based on gradient descent. However, since we introduce the local and global weights for parameters of the actor network, the update is treated differently from the normal actor update given as $\theta_{t+1}=\theta_t-\alpha_a\cdot(-\nabla_\theta J(\theta_t))$. During the weight update phase, the gradient of the loss with respect to the network's parameters is calculated for each local network, and also for the global network by taking into account the averaged contribution from all local networks. This setup is akin to an ensemble, where each local network learns its task, but also contributes to the learning of the global network. Each local weight is updated by extracting the local gradient weights $\theta_n$ from (11), and this portion of the weight is updated as

$$\theta_n = \theta_n + \alpha^{\theta}\cdot\nabla_{\boldsymbol{\theta_n}} J(\boldsymbol{\theta_n}), \tag{12}$$

where $\nabla_{\boldsymbol{\theta_n}} J(\boldsymbol{\theta_n})$ is the policy gradient loss with respect to the local network and $\alpha^{\theta}$ is the local actor's learning rate.

For the global weight, $\Theta^z$, connected to all local networks, the update procedure is to extract the weight from (11). Thus, the global weight is updated to have the average $\Theta^z$ as

$$\Theta^z = \Theta^z - \zeta\nabla_{\Theta^z}\left(\frac{1}{Z_s}\sum_{n=1}^{Z_s}(\nabla_{\boldsymbol{\theta_n}} J(\boldsymbol{\theta_n}) + \lambda \mathrm{L}^n_{\text{Critic}})\right), \tag{13}$$

where $\lambda$ is a weighting factor that balances the importance of the policy gradient loss and the value function loss in the global network's update, $\zeta$ is the learning rate for the global network, and $\mathrm{L}^n_{\text{Critic}}$ is value function loss for the critic network given by

$$\mathrm{L}^n_{\text{Critic}} = (V_{\Theta^p}(s_t) - (r_t + \gamma V_{\Theta^p}(s_{t+1})))^2. \tag{14}$$

This is to ensure that the global network learns to produce representations that are beneficial across board, enhancing the local networks' ability to make predictions and select actions that maximize rewards.

The critic network employed in this work is the same as that of the conventional actor-critic DRL.

*2) Problem Reformulation:* To apply the modified actor-critic DRL, we reformulate the cell switching problem as an MDP with the MBS serving as the agent.

*a) States:* The states of the modified actor-critic DRL comprises the traffic load of MBS and that of the SBSs under its coverage before cell switching and traffic offloading take place, as well as the current on/off status of the SBSs. Thus, the state can be defined as

$$s_t = [\psi_t^{i,2}, \psi_t^{i,3}, \psi_t^{i,4}, \ldots, \psi_t^{i,Z_b}, \Gamma_t^{i,2}, \Gamma_t^{i,3}, \ldots, \Gamma_t^{i,Z_b}]. \tag{15}$$

**Algorithm 2** Modified Actor-Critic DRL

**Require:** Learning rate Global Actor ($\alpha^{\Theta},\zeta^{\Theta} \in (0,1]$),
: Learning rate Local Actor ($\alpha^{\theta} == \alpha^{\Theta}, \in (0,1]$)
: Critic Learning ($\alpha^{\omega}$), discount factor ($\gamma$), $\forall \in (0,1]$
: Number of episodes $M$, Number of SBSs $Z_{\text{s}}$
: Number of time steps per episode $T$

1: **Input**: Initialize the actor network $\pi(a|s,[\Theta:\theta_i]:\forall i \in Z_{\text{s}})$
   Initialize the critic network $V(s,\omega)$
2: **for** $m=1$ to $M$ **do**
3: Reset the environment state $s_t$
4: **for** $t=1$ to $T$ **do**
5: Observe state $s_t$, and ($\forall\ i\ in\ Z_{\text{s}}$) action $a_{t_i} = \cup(\mu_{\Theta+\theta_i}(s_t))$, then $a_t = \cup a_{t_i} \forall\ i\ in\ Z_{\text{s}}$
6: Execute actions $a_t$ in the environment and observe the next state $s_{t+1}$ and reward $r_{t+1}$
7: Calculate $TD = r + \gamma V(s_{t+1},\omega_t) - V(s_t,\omega_t)$
8: **Accumulate and Update Critic Network**
   $\partial\omega = \nabla_{\omega}(TD)^2$ Accumulate
   $\omega = \omega - \alpha^{\omega}\partial\omega$ Update
9: **Accumulate Actor Network**
10: **for** $i=1$ to $Z_{\text{s}}$ **do**
   $\partial(\theta_i) = \nabla_{(\theta_i)} \log\pi_{(\theta_i)}(s_t,a_{t_i})(TD)$
11: **end for**
12: **Update Global Actor Network**
   Apply Equation (13)
13: **Update Local Actor Networks**
14: **for** $n=1$ to $Z_{\text{s}}$ **do**
   $\theta_n = \theta_n + \alpha^{\theta}\partial(\theta_n)$
15: **end for**
16: Update the current state $s_t = s_{t+1}$
17: **end for**
18: **end for**

*b) Actions:* The actions of the proposed modified actor-critic DRL comprise switching off/on the different types of SBSs in the network, such that

$$\begin{aligned} a_t &= \{\Gamma_t^{i,2},\Gamma_t^{i,3},\Gamma_t^{i,4},...,\Gamma_t^{i,Z_b}\} \\ &= \{\Gamma_t^{i,j} | j \in \{2,...,Z_b\}\}. \end{aligned} \tag{16}$$

Therefore, the complete action vector $a_t$ specifies the on/off decision for each SBSs at time $t$.

*c) Reward:* In this work, we design the reward so that the agent will explore different action combinations, and hence we set different rewards based on the condition of the network. Moreover, we normalize the reward to avoid a large network update to allow the network to learn parameters smoothly. The normalization factor for positive reward is the maximum energy saved that keeps updated every time $t$ and is given by

$$P_{\text{saved},t} = P_{\text{total}}(\psi_t^{i,j}) - P'_{\text{total}}\ (\psi_t^{i,j},\Gamma_t^{i,j}), \tag{17}$$

where $P'_{\text{total}}\ (\psi_t^{i,j},\Gamma_t^{i,j})$ is the total power consumption of the UDHN after cell switching operations is applied. Initially, the maximum power saved is set to $0$, and at every time step $t$ during training, when the power saved is greater than the previous maximum value, the maximum power saved is updated. The maximum power saved can be expressed as $P_{\text{saved}}^{\max} = \max_{t\in\{1,2,...,TM\}} P_{\text{saved},t}$, where $P_{\text{saved}}^{\max}$ is given as the largest power saved among all the episodes ($M$) and trajectory TM. And the reward at each time step is given as

$$r_t = \begin{cases} -1, & (\psi_t^{i,1} + \sum_{\substack{\Gamma_t^{i,n}: \\ \forall\Gamma_t^{i,n}=0}} \psi_t^{i,n}) > 1 || (P_{\text{saved},t} < 0) \\ -0.1, & \text{if no SBS is switched off} \\ \frac{P_{\text{saved},t}}{P_{\text{saved}}^{\max}}, & \text{if } P_{\text{saved},t} > 0. \end{cases} \tag{18}$$

Algorithm 2 shows the pseudo code for implementing the modified actor-critic DRL algorithm to determine the optimal cell switch off/on policy for energy optimization. The essential advantage of the proposed approach is that introducing both global and local weights in the actor enables effective exploration of a vast discrete joint action space. In addition, scaling to large state spaces is facilitated by deep neural network function approximators for the actor and critic.

### C. The Complete FAMAC Framework

The complete FAMAC framework is illustrated in Fig. 2 while the implementation procedure is presented in Fig 3. The procedure for implementing the proposed FAMAC cell switching framework involves 2 phases: offline and online.

*1) Offline Phase:* This is the phase where both the FL traffic prediction model and the modified actor-critic DRL algortihm are trained using the historical data from the network. The procedure of the offline training is discussed in the following:

*a) Network Initialization:* The first step is initializing the UDHN infrastructure. The MBS provides umbrella coverage and comprises several SBSs including RRHs, micro, pico, and femto that are deployed in hotspots for capacity boosting.

*b) Historical Data Collection:* With the network established, historical traffic load data are collected over time and stored locally at each SBS. This raw data remains at each SBS and is not shared.

*c) FL for Traffic Prediction:* An FL model is developed using the historical data collected at each SBSs for distributed privacy-preserving traffic prediction. The MBS is provided with knowledge of expected traffic loads of the SBSs in its coverage without needing to acquire actual traffic data, which leads to significant communication overhead (Algorithm 1).

*d) Modified Actor-Critic DRL Agent for BS Switching:* A customized actor-critic DRL agent is developed using the historical data collected from the network to learn intelligent policies for dynamic cell switching off that minimizes network power consumption over time (Algorithm 2).

*2) Online Phase:* This is the stage where the proposed FAMAC is installed in the network to provide real-time cell switching operation using the traffic load predicted by the trained FL model. The procedure is as follows:

*a) Real-time Traffic Prediction:* For real-time operation, at each time step, the trained FL model provides the MBS with predicted traffic loads for the SBSs under its control based on current conditions.

*b) Optimized Cell Switching:* The predicted traffic loads and the current on/off state of the SBSs are input as states to the trained modified actor-critic DRL agent. Based on this, using its learnt stochastic policy, the agent selects the optimal set of SBSs to switch off for the current traffic.

*c) Execution:* The joint switch off actions are executed in the network environment.

*d) Continuous Adaptation:* The principle of traffic prediction adopted when developing the FL model is a time series approach. This means that the previous traffic load of the network predicts the future traffic of the network. It is therefore essential to update the historical traffic database from which the future traffic load of the network is predicted by obtaining more recent data from the network within a given time interval $T$ (where $T$ could be once a day or a week, etc.). This would ensure that the modified actor-critic DRL agent keeps adapting its cell switching strategy to the more updated SBS traffic predictions obtained from the FL model.

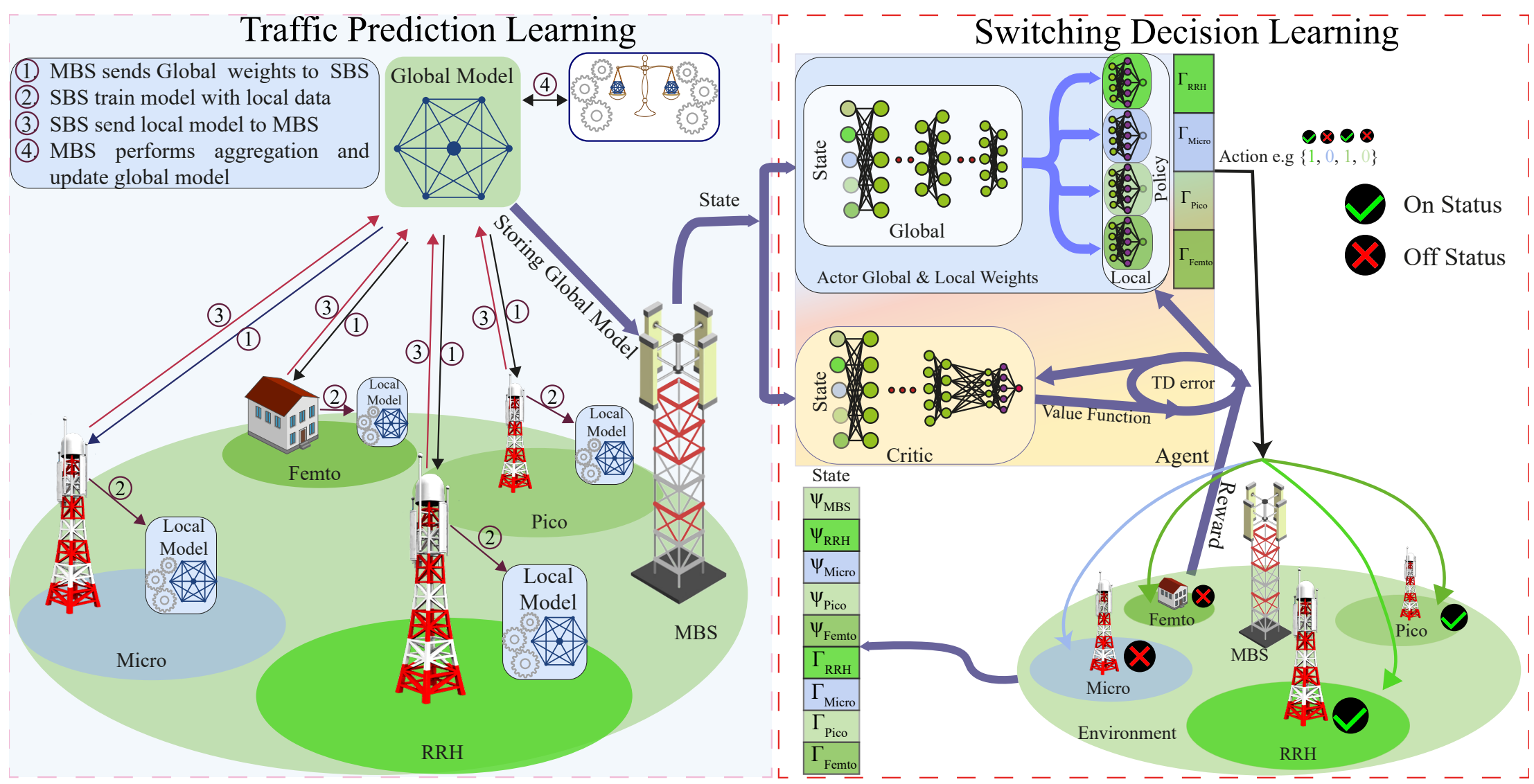

Fig. 2. The complete FAMAC framework comprising the FL-based traffic prediction and modified actor-critic DRL algorithm for BS switch off/on.

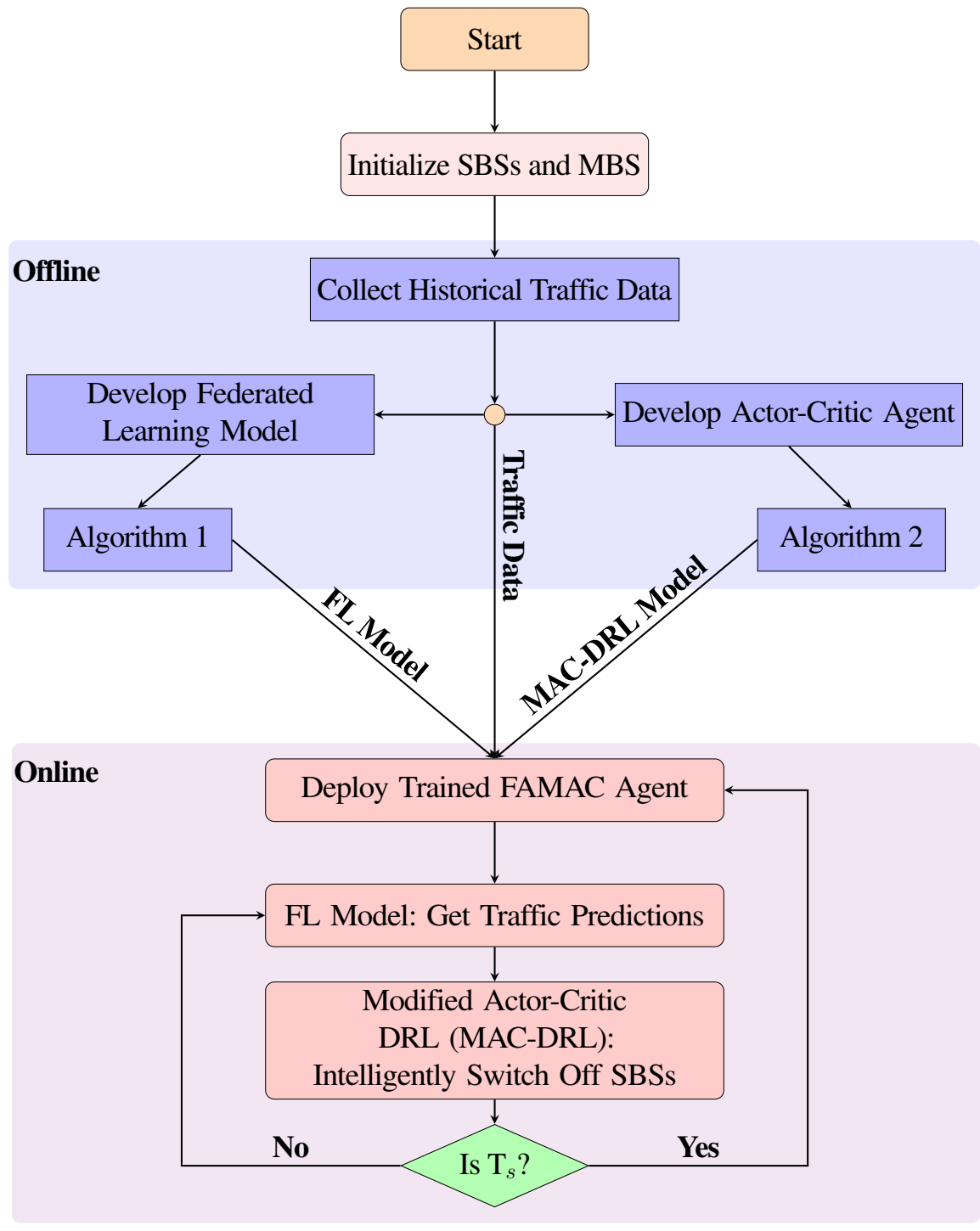

Fig. 3. The implementation procedure of the FAMAC framework.

In a nutshell, FAMAC combines FL for distributed traffic prediction and a modified actor-critic DRL for huge discrete action spaces to enable intelligent and secure data-driven optimization of cell switching for minimizing network power consumption across fluctuating traffic loads.

### *D. Dataset and Pre-Processing*

To obtain information about the traffic load of the network, a dataset comprising the call detail record (CDR) of the city of Milan [43] is utilized during the simulation, which comprises calls, text messages, and internet activities carried out in each grid every 10 minutes for two month. Then, the CDR of each grid is assumed to be the traffic load in this work since they represent the volume of network resources utilized by the users within each grid for each time slot. However, only the internet activity levels are considered as the traffic load for the BSs when processing the data because we are particularly interested in investigating 6G networks which is internet protocol (IP) based. Furthermore, to obtain the traffic load of an MBS, we combine the internet activity level of two randomly selected grids. In contrast, the internet activity of a single grid is considered to be the traffic load of each SBS. Lastly, we normalize the traffic loads of each BS with respect to the total amount of radio resources in each type of the BS in the UDHN (i.e. MBS, RRH, micro, pico and femto BSs). For the FL model development, the traffic load of 31 days, with each day comprising 144 samples, is obtained from the Milan dataset. 60% of the total samples (2680 samples) is used for training, 20% (892 samples) for validation, and the remaining 20% for testing (892 samples).

## VI. Performance Evaluation

We evaluate the performance of the proposed cell switching framework using various metrics and compare its performance with benchmark algorithms. It should be noted that the proposed framework is designed to be implemented at each MBS. The size of each MC is related to the number of SBSs included with the MC, i.e., $Z_s$. In this work, each cell is considered square-shaped with a dimension of $h$, where $h = 235$ meters, based on the structure of the real dataset used. Further details about the dataset is elaborated upon in Section V-D. To investigate the impact of $Z_s$ on cell switching performance, $Z_s$ is considered a variable in the simulation campaigns rather than a fixed value. Consequently, the size of the simulation environment is also variable. The area of the simulation environment, which is the accumulation of square-shaped grids given in the Milan dataset, is calculated as $A_s = Z_s h^2$. For a complete square area, where $Z_s = z_s^2$ with $z_s = 2z$ and $z \in \mathbb{Z}^+$, the dimension can be computed as $D_s = \sqrt{Z_s}$, where $D_s \in \mathbb{Z}^+$.

The power consumption parameters of the different types of BSs can be found in [35] (Table 2 therein), while the parameters used in the simulation of the FL model and modified actor-critic model are presented in Table II.

TABLE II
SIMULATION PARAMETERS

| Parameter | Value |
|---|---|
| **General** | |
| Number of time slots | 144 |
| Number of days | 1 |
| Number of grids per MBS | 2 |
| Number of grids per SBS | 1 |
| Bandwidth of MBS | 20 MHz |
| Bandwidth of RRH, Micro, Pico, Femto | 15, 10, 5, 3 (MHz) |
| **Traffic Prediciton LSTM** | |
| Dropout, Window Size | 0.2, 59 |
| LSTM Layers, Units, Bidirectional LSTM | 3, window size, Enabled |
| Dense Units, Activation, Batch Size | 1, Linear, 16 |
| Loss Function, | Root Mean Squared Error |
| Optimizer, Epoch | Adam, 100 |
| **Modified Actor-Critic** | |
| Optimizer/learning rate ($\alpha$,$\zeta$) | Adam (0.01, 0.015) |
| Global layers Actor | {32, 64, 128, 256} |
| Local layers Actor | {128, 64, 32} |
| Output layers Actor | $(Z_b - 1)$ - sigmoids |
| Critic layers | {32, 64, 128, 256, 512, 32} |
| Output Critic layer | Linear |
| Exploration rate | 0.01 |
| Discount Factor | 0.90 |

### A. Benchmarks

*1) All-Always-ON (AAO):* No cell switching is implemented, thereby all the BSs are constantly kept on. Thus, no energy saving is obtained and QoS is always maintained.

*2) Exhaustive Search (ES):* This method is always guaranteed to find the optimal policy as it searches through all the possible combinations to determine the set of SBSs to switch off. This method also ensures that QoS of the network is always satisfied by considering the constraint in (8).

*3) Multi-Level Clustering (MLC):* The MLC approach is a clustering-based BS switching framework based on the $k$-means algorithm, wherein the SBSs deployed in the coverage area of the MBS are repeatedly clustered according to their traffic load and the cluster(s) of SBSs with the least traffic load are switched off. The cluster giving the minimum power consumption is selected as the optimum cluster.

*4) Threshold-based Hybrid cEll SwitchIng Scheme (THESIS):* THESIS is an improvement on the MLC algorithm. It is a threshold-based hybrid algorithm comprising the $k$-means clustering and ES algorithms [4].

These algorithms represent a wide spectrum of cell switching strategies, ranging from the most basic to the most advanced. AAO serves as the baseline approach, where no cell switching is implemented, providing an upper bound on energy consumption. Conversely, ES represents the optimal cell switching strategy but becomes computationally infeasible for large-scale networks, despite offering a lower bound on energy consumption. MLC and THESIS are two state-of-the-art algorithms that aim to balance energy efficiency and computational complexity. MLC employs a clustering approach to reduce the search space and computational overhead, while THESIS combines clustering with ES to improve energy savings while maintaining lower computational complexity than ES. The inclusion of these algorithms in the performance evaluation allows for a comprehensive comparison of FAMAC with specifically designed cell switching methods for UDHNs.

### B. Performance Metrics

*1) Root Mean Square Error (RMSE):* RMSE is the cost function employed to evaluate the error of the bi-directional LSTM model. It depicts the discrepancy between predicted and actual values using Euclidean distance.

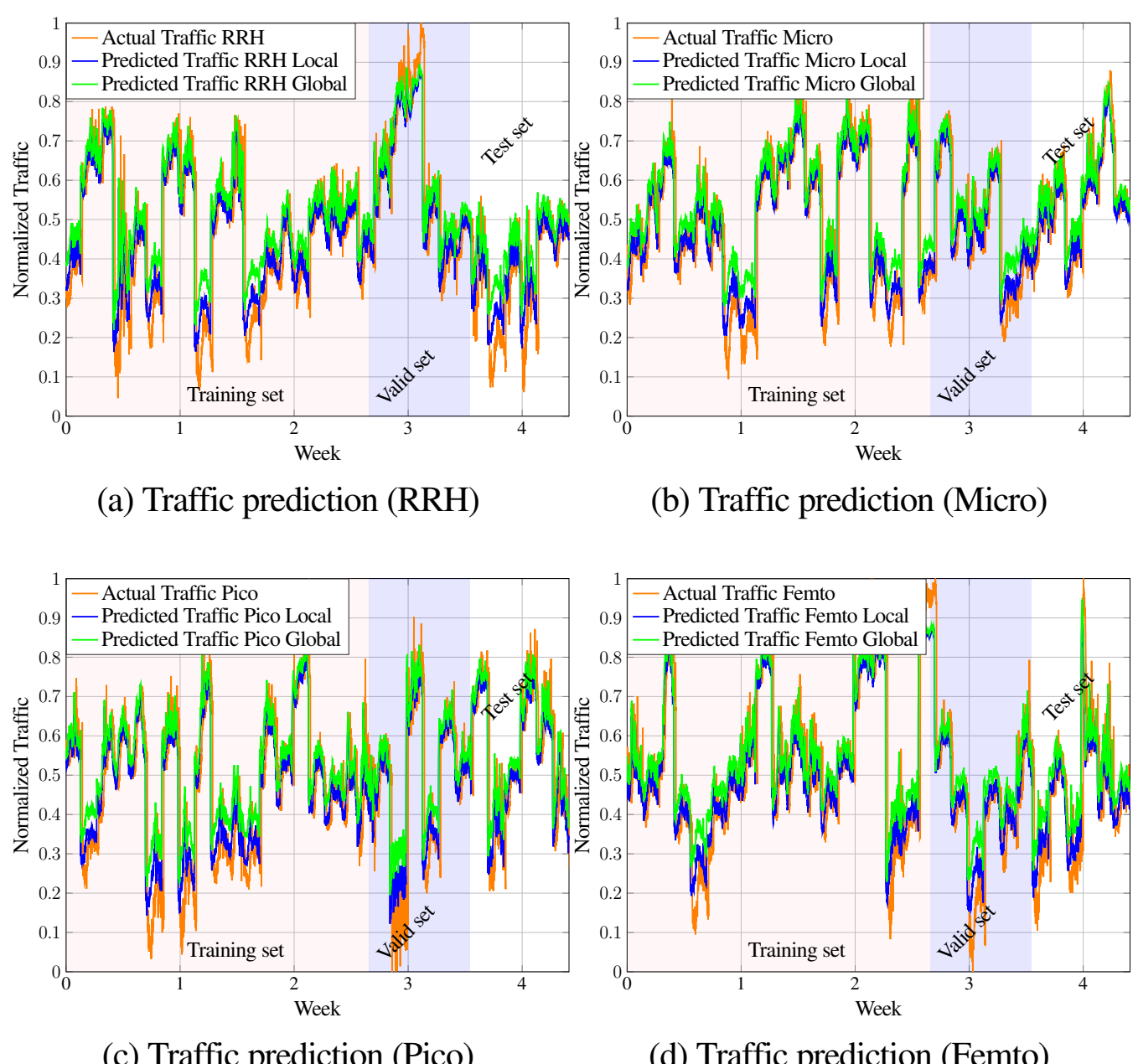

(a) Traffic prediction (RRH) (b) Traffic prediction (Micro)

(c) Traffic prediction (Pico) (d) Traffic prediction (Femto)

Fig. 4. Performance of the proposed FL based traffic prediction.

TABLE III
COMPARISON OF RMSE OF LOCAL AND GLOBAL MODEL OF EACH SBS TYPE

| SBS Type | Type of Model | RMSE |
|---|---|---|
| RRH | Global | 0.1221 |
| | Local | 0.0987 |
| Micro | Global | 0.1101 |
| | Local | 0.0902 |
| Pico | Global | 0.1318 |
| | Local | 0.1145 |
| Femto | Global | 0.1261 |
| | Local | 0.1080 |

*2) Power Consumption:* This is the instantaneous power consumption of the network (in Watts) obtained during simulations using (3) for the proposed and benchmark approaches.

*3) Energy Saved:* This metric quantifies the total amount of energy saved in the network (in Joules) when the switching methods are applied for a certain duration. It is computed against the case when no switching is applied (i.e., AAO).

*4) Coverage loss ($C_l$):* It is a measure of the percentage difference in the total volume of traffic that is supported by the network before and after the BS switching is performed.

### C. Results and Discussions

In this section, we evaluate the performance of the FAMAC framework comprising the FL-based traffic prediction model and the modified actor-critic DRL-based cell switching model.

*1) **Comparison of Traffic Prediction Techniques**:* Fig. 4 presents the traffic prediction performance of the local and global FL models, while Table III depicts the RMSE values obtained during model training for each type of SBS. For RRHs, we observe that the local model demonstrates better prediction accuracy than the global FL model, as evidenced by its lesser RMSE of 0.0987 compared to 0.1221 for the global model (Table III). When the prediction plots in Fig. 4a are considered, the local model tracks the fluctuations in the actual RRH traffic much more closely, especially in the initial training weeks. For example, in Week 1, the local model accurately captures a sharp rise from 0.1 to 0.8 normalized load, while the global model underestimates the surge. This is because the local model can learn unique traffic characteristics and trends specific to each RRH site. In contrast, by aggregating data across locations, the global model generalizes and misses some location-specific variations. However, the global model can identify overall network-wide changes and emerging patterns in the RRH traffic (like broader increases or decreases) that local models may overlook.

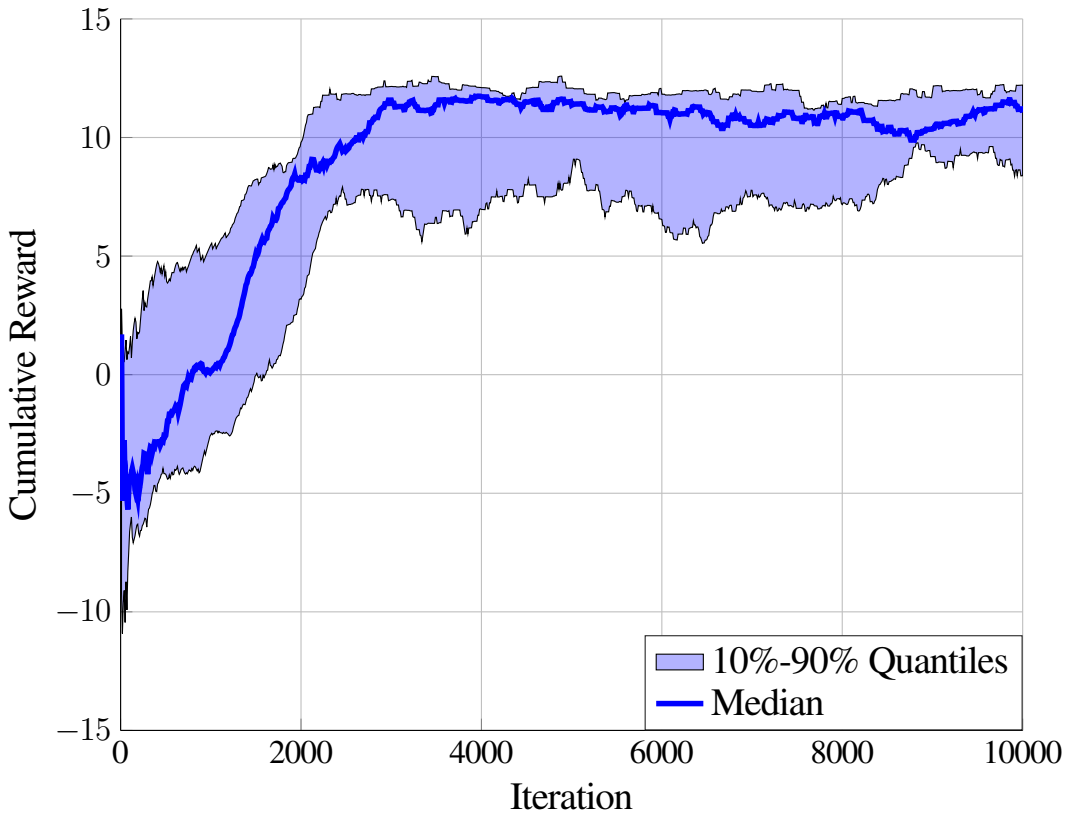

Fig. 5. Proof of Convergence curve. The darker line shows the median over seeds. We average the two extreme values to obtain the shaded area (i.e., 10% and 90% quantiles with linear interpolation).

For micro SBSs, again the local model attains lower prediction error (RMSE 0.0902) than the global federated model (RMSE 0.1101) as shown in Table III. Also, from Fig. 4b, the local model consistently anticipates the ups and downs in the real micro SBS traffic better across training and testing weeks. This highlights its ability to closely learn distinct micro SBS traffic behaviors that is unique to each specific location. Albeit being less accurate, the global model provides advantages like privacy preservation, reduced communication overhead, adaptability to new locations, and identifying shared micro SBS traffic patterns across the network.

Regarding pico SBSs, the global and local models achieve very similar RMSE performance of 0.1318 and 0.1145, respectively, as shown in Table III. Examining the prediction plots in Fig 4c, both modeling approaches track the pico SBS traffic fluctuations reasonably well. This suggests that pico SBS traffic may exhibit more homogeneous characteristics across locations. Hence, for pico SBSs, the global and local models are comparable in prediction accuracy, while the global model provides benefits like privacy and reduced signaling overhead through FL. Lastly, for femto SBSs, the global and local models attain close RMSE values of 0.1261 and 0.1080, respectively (Table III). As visualized in Fig. 4d, their predictions equally capture the variations in the actual femto SBS traffic over the weeks. This indicates that femto SBSs experience consistent localized traffic patterns, making a specialized local model unnecessary. The global model performs on par for femto SBSs with uniform traffic, while still providing the generalization and communication efficiency benefits of FL. Overall, even though the local models achieve relatively lower prediction error compared to the global model, the global FL model provides advantages like privacy, reduced signalling overhead, and adaptability.

*2) **Proof of Convergence**:* Fig. 5 illustrates the convergence behavior of the proposed modified actor-critic DRL algorithm during training. It demonstrates how the proposed modified actor-critic DRL algorithm learns the optimal cell switching policy to minimize the total network power consumption using the reward function in (18). A key aspect is penalizing the agent for violating the MBS traffic limit or achieving no power savings.

In the initial phase (between iteration 1 and 1000), high volatility in median reward occurs as the agent explores the actions, frequently violating constraints (i.e., negative rewards). The wide percentile spread shows high variance across seeds as the agent struggles to find useful switch off patterns. In the middle stage (between iteration 1000 and 3000), the volatility reduces as the agent learns to satisfy constraints and achieve modest energy savings, receiving slightly positive rewards. The interquartile range narrows as the performance becomes more consistent across seeds. In the final phase (from iteration 3000), the median reward increases and stabilizes close to 10 as the agent converges on optimized policies, achieving strong positive rewards. The tight percentile spread proves the reliability of this solution across training conditions. The agent learns effective switch off strategies tailored to traffic loads and BS types that maximize cumulative reward.

The upward trend of the median reward demonstrates that the agent successfully learns to improve its policy over episodes by interacting with the environment. The feedback on power consumption and coverage loss, formalized in the reward function in (18), enables the model to adapt its decisions to optimize the target of minimum network power. Violating constraints leads to negative rewards that teach the agent to prioritize feasibility. The stabilization of high median reward and narrow percentiles proves the agent converges on an optimized solution that is robust to variability in training.

*3) **Comparison of Instantaneous Power Consumption**:* Fig. 6a shows the instantaneous power consumption of the network over a whole day for 20 SBSs, while Fig. 6b and Fig. 6c illustrate the distribution of the instantaneous power consumption during on-peak and off-peak periods when the switching methods are implemented. First, as observed from Fig. 6a that the instantaneous power consumption varies with the traffic load of the network over the 24-hr period when FAMAC and the benchmark methods are applied. However, compared to AAO, the switching methods are able to reduce the power consumption of the network during both low and high traffic periods even though the magnitude of reduction in network power consumption is much higher during off-peak hours because of the opportunities to switch off more SBSs.

Second, a closer look at the on-peak and off-peak power consumption analysis, as shown by the box plots in Fig. 6b and Fig. 6c, reveals the superior performance of FAMAC compared to both THESIS and MLC because its median, minimum, and maximum power consumption values are lesser during both on-peak and off-peak traffic periods, even though the power consumption margin is higher at off-peak periods. The justification for the higher performance of FAMAC during the off-peak periods is that the traffic load of the network is low during off-peak periods and there is more offloading capacity at the MBS and more SBSs are available to switch off. However, at high traffic periods, the performance of FAMAC and the benchmark methods become similar because there is less offloading capacity in the MBS, and most SBSs are highly

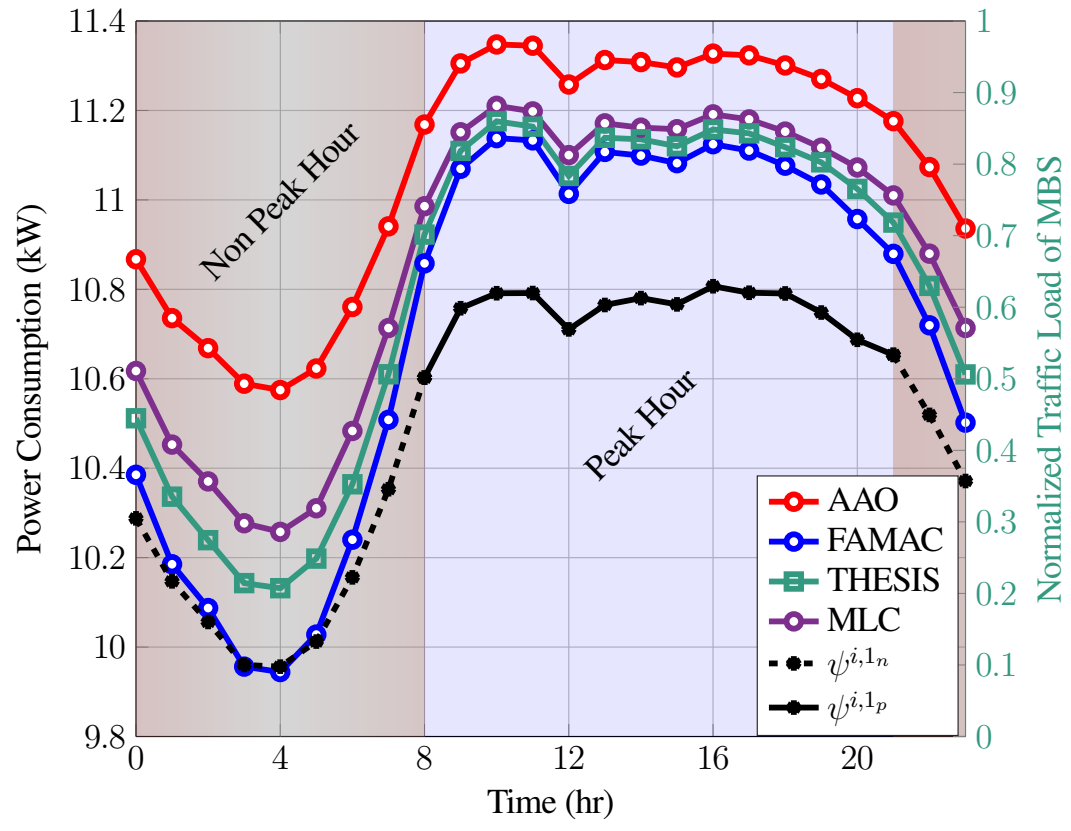

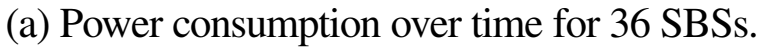
(a) Power consumption over time for 36 SBSs.

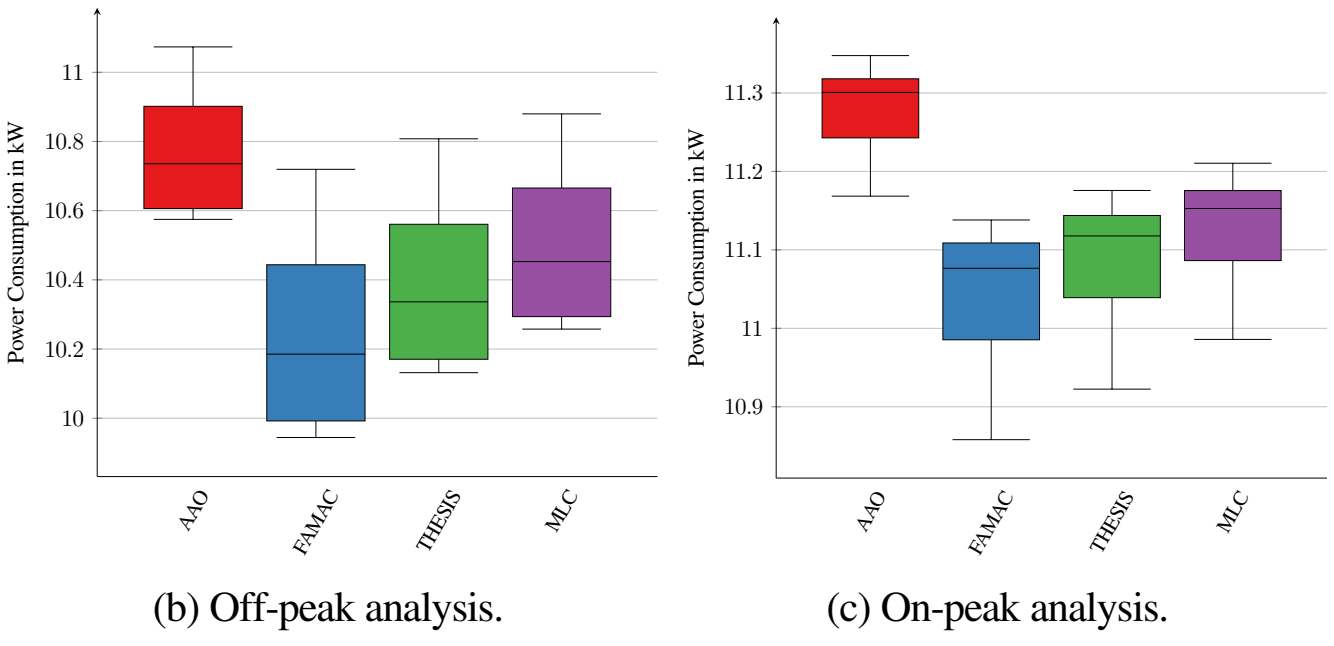

(b) Off-peak analysis. (c) On-peak analysis.

Fig. 6. Instantaneous power consumption over a 24-hr period for 36 SBSs.

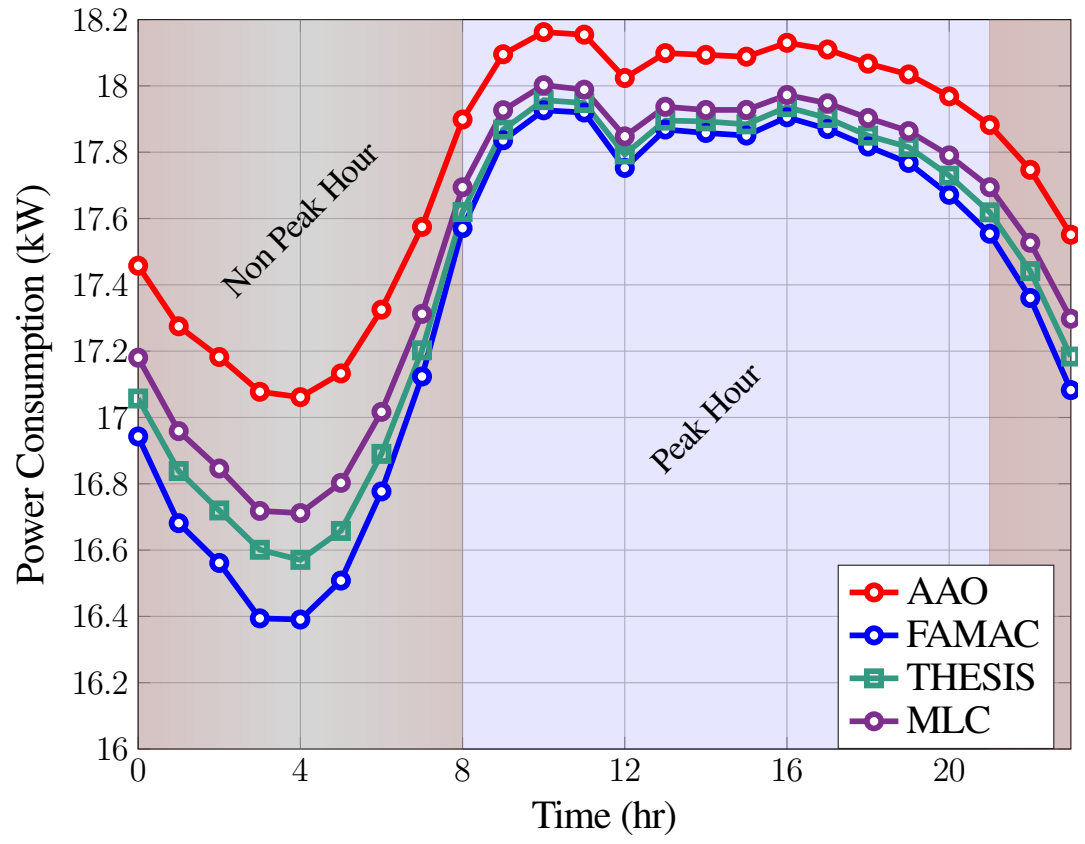

(a) Power consumption over time for 60 SBSs.

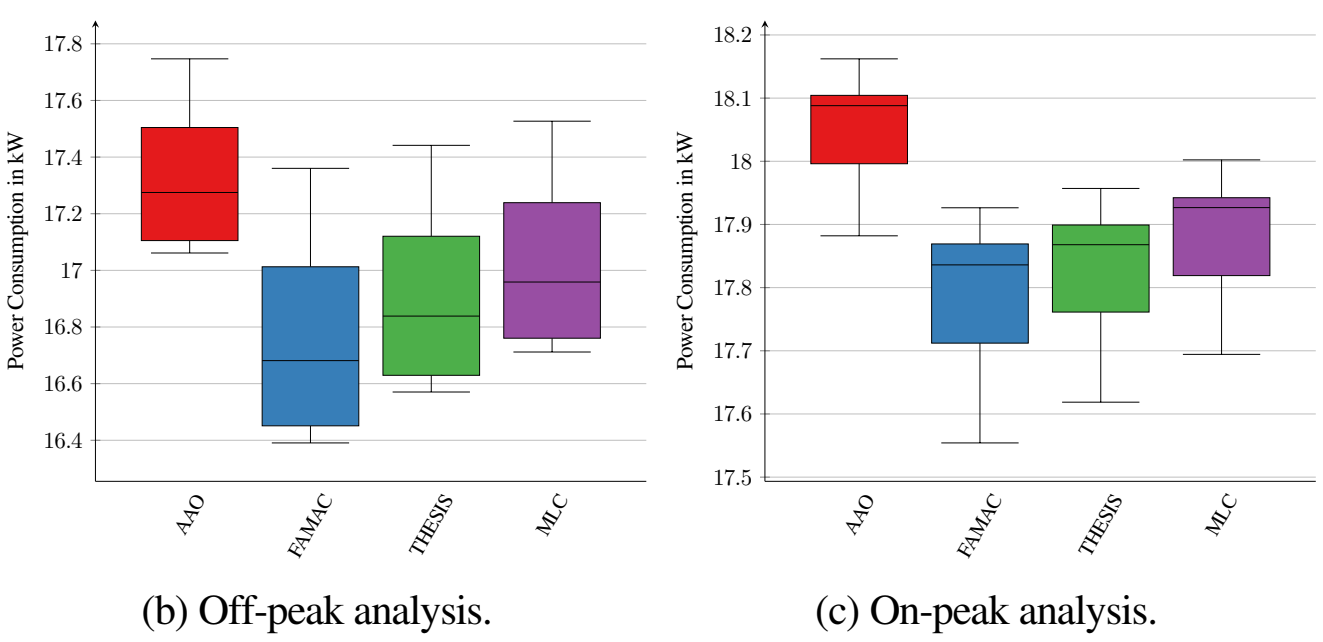

(b) Off-peak analysis. (c) On-peak analysis.

Fig. 7. Instantaneous power consumption over a 24-hr period for 60 SBSs.

loaded, so the switching options that FAMAC can select without violating the QoS constraint become very few thereby making its performance almost the same as the benchmarks.

Third, from Fig. 6a, Fig. 6b, and Fig. 6c, the performance of THESIS is next to that of FAMAC during both on-peak (Fig. 6a in legend $\psi^{i,1_p}$) and off-peak (Fig. 6a in legend $\psi^{i,1_n}$) periods. However, we still observe a wider margin in power consumption during low traffic periods compared to high periods. The reason for this discrepancy is that THESIS employs both clustering and ES methods which limits the amount of search spaces that it can explore to determine the optimal combination of SBSs to switch off when the traffic is low, but at high traffic periods, the larger search space that FAMAC has compared to THESIS does not offer much advantage due to the lack of sufficient offloading capacity at the MBS, accounting for the narrow margin at high traffic.

Fourth, it can be seen from from Fig. 6a, Fig. 6b, and Fig. 6c that asides AAO, which has a lesser performance than MLC, all other methods have superior performance both during on-peak and off-peak periods. The rationale behind the low performance of MLC is that it does not discriminate among the different types of SBSs when deciding which SBSs to switch off and offload their traffic to the MBS. Rather, it is more concerned about switching a cluster with the highest number of SBSs whose traffic load can be accommodated by the MBS. By so doing, it places more priority on the number of SBSs that can be switched off instead of the type of SBS without realizing that it is possible to consume less power when few SBSs with higher power consumption are switched off compared to when many SBSs with lesser power consumption are switched off. Hence, we observe a wide margin between MLC and THESIS and a much wider margin between MLC and FAMAC during low traffic periods. However, at high-traffic periods, the margin becomes very slim because the chances of switching off SBSs are very few due to being heavily loaded, and there will also be very little offloading capacity in the MBS to accommodate their traffic load, so the uniqueness of each algorithm is barely noticed.

Fig. 7a illustrates the instantaneous power consumption of the UDHN with 60 SBSs for a 24-hr period while Fig. 7b and Fig. 7c analyze the off-peak and on-peak power consumption of the UDN using box plots. First, we observe a similar trend in power consumption of the network in Fig. 7a as in Fig. 6 where 36 SBSs are deployed, except that the power consumption values are of higher magnitude compared to that of Fig. 6. This is because more SBSs are deployed in this scenario copared to the previous scenario, thereby resulting in an increase in the overall power consumption of the network. As with Fig. 6, the performance of FAMAC supersedes that of THESIS and MLC more significantly during periods of low traffic than high traffic because of the enhanced ability to determine the best policy in terms of the set of SBSs to select that would result in minimum energy consumption. The on-peak and off-peak power consumption analyses with 60 SBSs as illustrated in Fig. 7b and Fig. 7c also follow a similar trend as with Fig. 6b and Fig. 6c; FAMAC has a lesser median, minimum, and maximum power consumption values during on-peak and off-peak periods.

*4) **Comparison of Energy Saving**:* Fig. 8 shows the variations in the amount of energy saving that can be obtained when the number of SBSs increases from 8 to 120 for a 24-hr period. We can observe an increase in the amount of energy saving though with different magnitudes for all the methods between 8 and 60 SBSs, but after 60 SBS there is a decline in the magnitude of energy savings or almost steady energy savings irrespective of the increase in the number of SBSs. The rationale behind this observation is that as the number of SBSs begins to increase, there would be more opportunities to switch off more SBSs, and as a result, the energy savings begin to increase with high magnitude from 8-60 SBSs; however, from 60 SBSs upwards, the MBS becomes saturated such that the number of SBSs it can accommodate almost becomes constant. As such, the switching algorithms begin to select almost the same number and type of SBSs to switch off, making the magnitude of energy savings with successive SBS deployment to become very small.

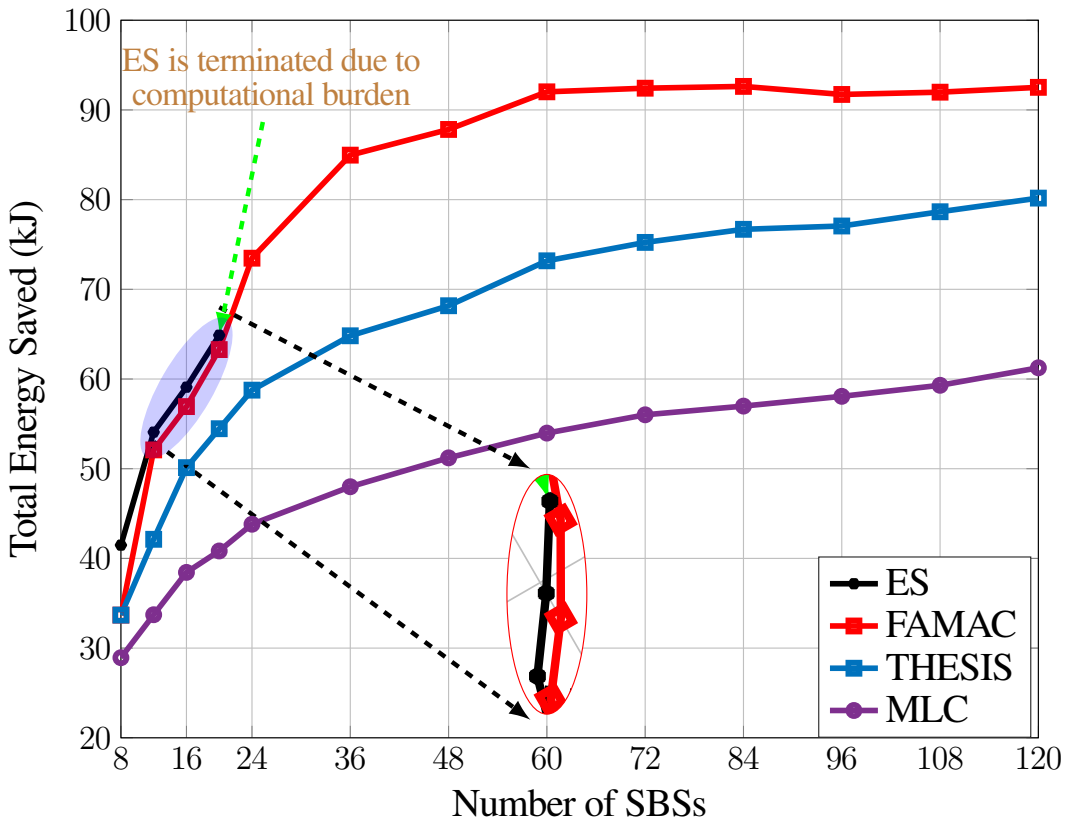

Fig. 8. Total energy saved in the UDHN over a 24-hr period.

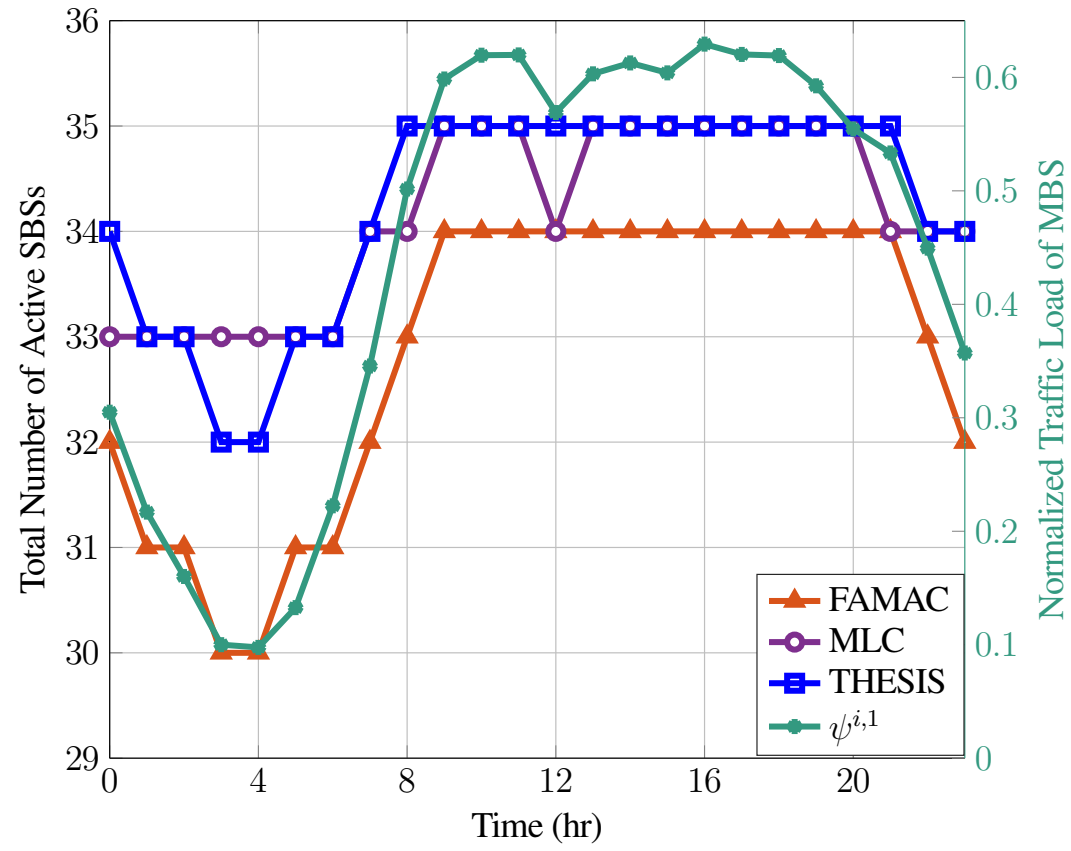

Fig. 9. The number of active SBSs after the switching off implementation.

Moreover, the ES algorithm has the best performance, because it is able to sequentially search through all the switching combinations to decide on the optimal number of SBSs to switch of per time, which would result in the maximum energy saving for the network. However, when the search space becomes very large, it becomes infeasible to sequentially search all the switching combinations to determine the optimal switching combinations because of the huge computational overhead that is involved, hence it is only feasible for small-sized networks. For this reason, we truncate the number of SBSs for ES to 20 in this work. Furthermore, we can see that even though the performance of FAMAC is quite lower than that of ES at 8 SBSs, from 12 to 20 SBSs, we observe an improvement in its performance—being equal to that of ES at 20 SBS. This clearly shows that FAMAC is a very close approximation of ES and a more suitable alternative because of its scalability to the higher number of SBSs deployment with lesser computational overhead.

In addition, from Fig. 8 we observe that FAMAC outperforms both THESIS and MLC benchmark methods in terms of energy saving. This is because it is able to take advantage of the available offloading capacity in the MBS to switch off the higher numbers and higher power-consuming type of SBSs before it saturates at about 60 SBSs. That is why we observe a much stepper increase in energy savings from 16 SBSs until about 60 SBSs where the energy savings becomes almost constant. Increasing the number of SBSs beyond 60 does not translate to much additional energy savings because the MBS has limited capacity to accommodate more traffic being offloaded from a large number of SBSs. Therefore, FAMAC selects mostly the same SBSs to switch off after the number of SBSs reaches 60, providing little extra energy reduction. Although being very close, the energy saving of THESIS increases with a relatively lower magnitude compared to that of FAMAC. This is because THESIS employs both clustering and ES approaches to decide the optimal switching solutions and it is not able to consider all the SBSs in the network at once, limiting the optimality of the final switching policy that it produces. However, its performance is still better than MLC which has the worse energy saving performance because, unlike MLC, which tries to switch the optimal SBS cluster in the UDHN, THESIS tries to find the optimal combinations of SBSs within each cluster to switch off. Thus, THESIS is able to select a few SBSs with higher energy consumption to switch off compared to MLC, which is more concerned with switching off clusters with more SBSs.

Lastly, we also infer from Fig. 8 that the deployment of more SBSs does not always translate to more SBSs being switched off due to the capacity constraints that is imposed on the network. This is to ensure that the QoS of the network is also satisfied during cell switching. Hence, increasing the number of SBS would not always lead to an increased opportunity to switch off more SBSs which limits the amount of energy savings that can be obtained unless a provision is made to increase the offloading capacity of the MBS. Overall, the highest margin in energy saving between the proposed and the benchmark methods occur when 36 SBSs are deployed where the energy saving performance of FAMAC becomes 77% and 31.1% higher than THESIS and MLC, respectively.

*5) **Comparison of Number of Active SBSs**:* Three major observations can be derived from Fig. 9. First, the proposed method is specially designed to switch off a unique combination of SBSs, comprising more power-consuming SBSs, than both MLC and THESIS and, as a result, is able to save more energy. Second, the fact that an algorithm switches off more SBSs than another does not automatically translate to saving more energy as can be seen when we compare Fig. 6 and Fig. 9 where even though MLC has lesser active SBSs compared to THESIS, its power consumption is still higher than that of THESIS. The reason for this is that four different types of SBSs with varying power consumption values are deployed. Even though MLC switches off more SBSs from 3hr-4hr and at 12hr, THESIS switches off higher-power SBS types. Hence, despite switching off fewer SBSs, THESIS achieves superior total power savings over MLC. However, FAMAC has superior performance than MLC and THESIS because it is carefully designed to select not just the optimal number of SBSs but the type of SBS to switch off per time that would result in maximum energy saving in the network.

Third, the number of SBSs that can be switched off at any given time depends on the available capacity at the MBS as the traffic of SBSs that are switched off must be transferred to the MBS in order to maintain the QoS. This restricts the number of SBSs that can be switched off, placing a limit on the amount of energy that can be saved. That is why we observe that when the traffic load of the MBS is very high from 10hr to 20hr, a maximum of 2 SBSs can be switched off as compared to low traffic period between 3hr and 4hr where up to 6 SBSs can be switched off. This is because we design the BS switching problem in such a way that the QoS of users is given higher priority compared to the energy efficiency. This is consistent with practical networks as mobile network operators (MNOs) place more emphasis in providing quality service to their customers so as to drive up their revenue compared to switching off more SBSs to save energy.

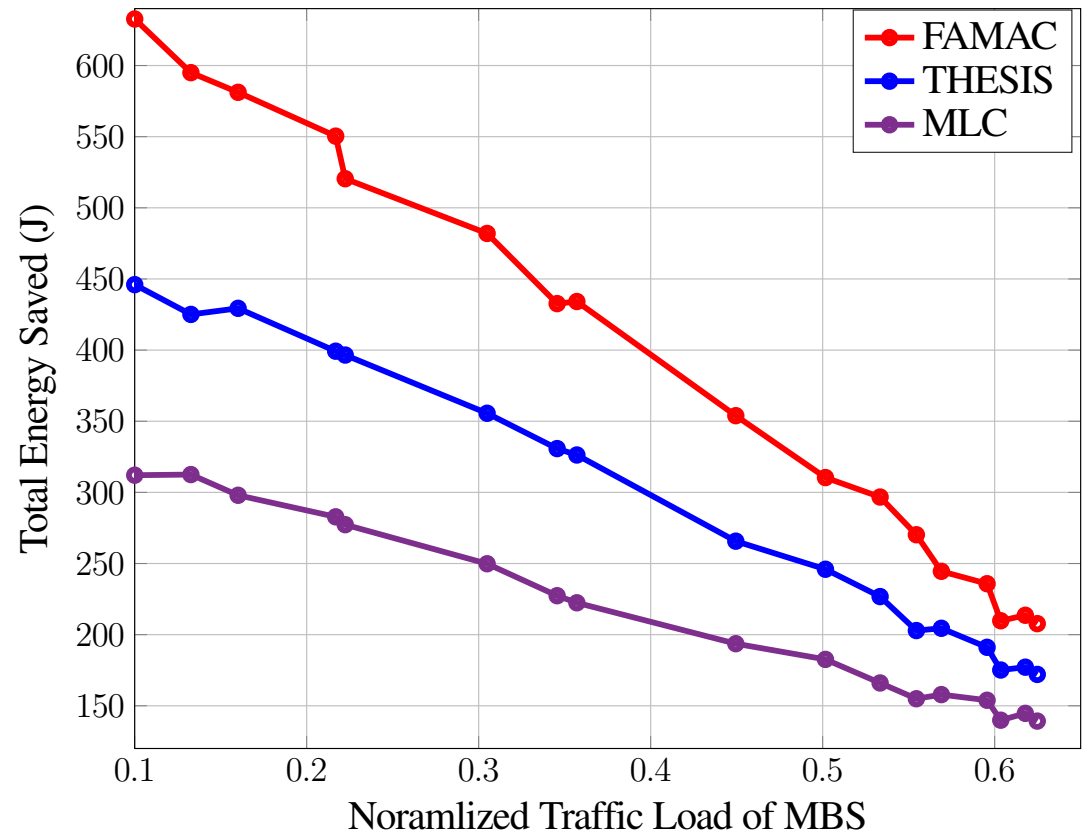

Fig. 10. The effect of the MBS traffic on the energy savings of the UDHN.

*6) **Impact of MBS Traffic on Energy Savings**:* Fig. 10 illustrates the amount of energy savings that can be obtained when the traffic load of the MBS varies from low to high. Here, we study the effect of the capacity constraints, that have been put in place to ensure that the QoS of the network is not violated, on the amount of energy savings. The energy saving of both the proposed and benchmark methods almost scales linearly with the amount of traffic demand on the MBS which means that when the MBS is lightly loaded, more energy can be saved in the network as it can accommodate more traffic load thereby allowing more SBSs to be switched off as opposed to highly loaded periods which the energy savings is very low because it can only accommodate little traffic and as such only very few SBSs can be switched off.

In addition, we can see that FAMAC outperforms both THESIS and MLC at both low and high traffic periods with energy savings of up to 99% and 42.3% greater than that of MLC and THESIS, respectively, during low traffic periods. The superior performance of FAMAC is because of its ability to select the optimal number and type of SBSs to switch off which will lead to maximum energy saving in the network without violating the capacity constraints of the MBS. We can also see here that THESIS performs better than MLC with an energy saving that is 40% higher than that of MLC—even though MLC can switch off more SBSs than it at some time instances as observed in Fig. 9—because it has the ability to discriminate among the various types of SBSs and select the few SBSs that would save a considerable amount of energy.

*7) **Analysis of SBS Power Consumption**:* Next, we present an analysis of the instantaneous power consumption breakdown of the various types of BSs that are deployed in the network using box plot in Fig. 11 for both the proposed and benchmark methods to further highlight the excellent performance of the proposed approach. The boxplot demonstrates the median (mean), the minimum, and maximum instantaneous power consumption values, and the skewness of the power consumption values which describes how the values are distributed across the mean value.

A seen in Fig. 11a, FAMAC has the lowest mean power consumption compared to THESIS and MLC. In addition, it has the lowest minimum and maximum power consumption values, and the power consumption values are evenly distributed about the mean compared to both THESIS and MLC where most of the power consumption values of RRH lie above the median mark. This indicates FAMAC prioritizes switching off more RRHs in the network than the other benchmarks.

We also observe in Fig. 11b that FAMAC still has a lower mean power consumption compared to THESIS and MLC. As for the maximum value, we see that all algorithms have almost the same maximum value while, in terms of minimum value, THESIS achieved the lowest. With respect to data skewness, the power consumption values of FAMAC are evenly distributed around the mean while most of the power consumption value of THESIS and MLC lie above the median value. From these observations, we see that the performance of FAMAC is still slightly better in terms of prioritizing the switching off of more micro SBSs. Further, it is observed in Fig. 11c that MLC has a slightly lower median power consumption compared to FAMAC, while THESIS has a higher median power consumption than both FAMAC and MLC. In terms of minimum and maximum power consumption, both MLC and FAMAC exhibit the same minimum power consumption, with MLC's maximum value slightly lower than that of THESIS. In addition, THESIS has the same maximum value as FAMAC, but its minimum value is much higher than MLC and THESIS. For power consumption skewness, most of the power consumption values of THESIS and MLC are lower than the median market, while that of THESIS is evenly distributed around the median. These observations suggest that MLC gives more priority to switching off pico SBSs, followed by FAMAC, while THESIS seems not to switch off any pico SBS as its median, maximum, and minimum power consumption values are almost the same as that of AAO. In Fig. 11d, we see that FAMAC has a lower median and minimum power consumption compared to THESIS and MLC, even though the proposed and benchmark methods have the same maximum value. The performance of MLC is almost the same as that of THESIS, with slightly lesser values. Here again, we see that THESIS has the same performance as AAO.

Fig. 11e showcases the median, minimum, and maximum power consumption values of the MBS. The power consumption value of the MBS with AAO is the lowest while FAMAC has the highest power consumption. The rationale behind this is that, with FAMAC, more SBSs can be switched off (as we see in Fig. 9), which means that more traffic is offloaded from SBSs to the MBS, and this translates to higher energy consumption in the network. Moreover, as shown in Figs. 11a to 11d, THESIS switches off more high-power RRH and micro SBSs compared to MLC. Even though THESIS switches off fewer total SBSs than MLC (Fig. 10), its shutdown of those specific power-hungry RRH and micro SBSs enables more traffic offloading to the MBS. This greater MBS offloading causes THESIS to consume more power overall compared to MLC. MLC has the lowest power consumption in MBS because it prioritizes switching off more pico and femto SBS, which are off lesser capacity and power consumption than micro and RRH, as a result, the amount of offloaded traffic is lesser than that of THESIS and FAMAC which also leads to a reduced energy consumption in the network.

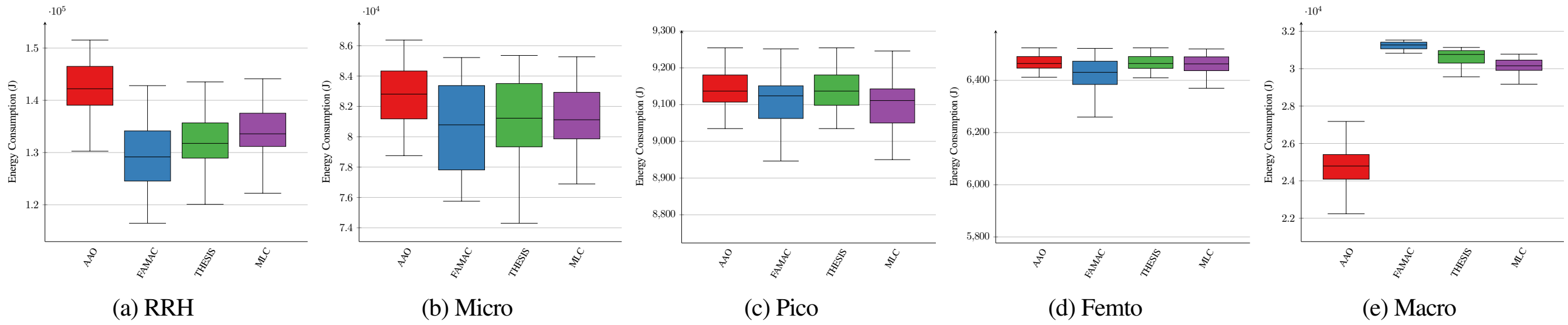

Fig. 11. Comparison of the energy consumption breakdown of the different types of BSs in the network for 36 SBSs.

*8) **Analysis of Switch Off Occurrence among SBSs**:* We further demonstrate the superior performance of the proposed framework using a polar plot as illustrated in Fig. 12, which compares the average daily switch off occurrences for the different SBS types. The plot shows the results for a network with 36 SBSs (9 SBSs of each type). The polar plot has four quarters, each representing an SBS type (i.e., 1-9 RRHs, 10-18 micro SBSs, 19-27 pico SBSs, and 28-36 femto SBSs). Within each quarter, there are radii, each 10 degrees apart, which denote the number of each type of SBS, and concentric circles, which indicate the average frequency of switching off of each SBS type when FAMAC, THESIS, and MLC are implemented.

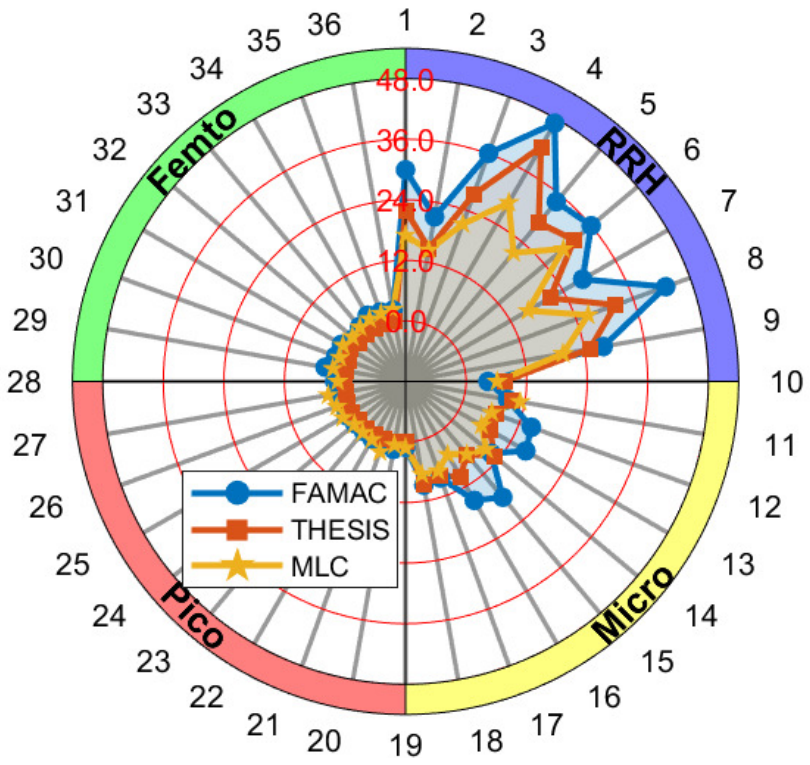

Fig. 12. Polar plot illustrating the average daily switch off occurrences for each BS type using proposed and benchmark methods.

We see that FAMAC switches off around 7 out of 9 RRHs, 5 out of 9 micro SBSs. For pico and femto SBSs, only 2-3 out of 9 are switched off on average. Thus, FAMAC preferentially switches off more RRHs and micro SBSs compared to pico and femto SBSs. By targeting high power consumption SBSs, FAMAC is able to maximize energy savings since RRH and micro SBSs have higher power consumption compared to pico and femto. The polar plot also indicates that THESIS is able to switch off a moderate number of RRHs and micro SBSs; however, it rarely switches off pico and femto SBSs. This is consistent with our analysis so far that THESIS focuses on selective switching off of high-power SBSs because its hybrid approach prevents it from exploring the full search space, thereby limiting its overall energy savings. Fig. 12 also shows that MLC has a more balanced switching across RRH, micro, pico, and femto SBSs, but the number of SBSs switched off is lower overall compared to FAMAC. However, because it does not target the switching off of more SBSs with high power consumption, it is not able to maximize power reduction, thereby resulting in its lowest performance.

*9) **Coverage Loss Analysis**:* It should be noted that due to the way the cell switching problem is formulated in this work, both the proposed and benchmark methods are designed in such a way that we pay more stringent attention to the QoS constraint given in (8) during simulations by ensuring that the traffic of users associated with SBSs that are to be switched off is always re-associated with the MBS before switching them off. As a result, the coverage loss of FAMAC, THESIS, and MLC are the same (i.e., the coverage loss obtained from all methods is zero, which is the case where the traffic demand before and after cell switching is implemented is the same). We adopt this approach when designing the proposed and benchmark methods because we understand that MNOs emphasise satisfying user service demands in practical networks rather than sacrificing user QoS to save energy.

TABLE IV
NUMBER OF SBSS AND INFERENCE TIME

| No. of SBSs ($Z_s$) | 4 | 8 | 12 | 16 | 20 | 24 | 28 | 32 | 36 |
|---|---|---|---|---|---|---|---|---|---|
| Time (s) | 0.009 | 0.015 | 0.026 | 0.035 | 0.040 | 0.048 | 0.055 | 0.063 | 0.074 |
| Action Space | 4 | 8 | 12 | 16 | 20 | 24 | 28 | 32 | 36 |

*10) **Actions and Time Computation Analysis**:* The analysis of both time complexity and action space is illustrated in Table IV. The action space demonstrates a linear increase as the number of SBSs rises. This is because each additional SBS contributes proportionally to the action space. Moreover, the time complexity for inference exhibits a linear growth, following the Big-O notation $O(n)$. This linear increase implies that as the number of SBSs increases, the computational resources required for inference grow proportionally. This enhancement results in a significantly faster solution compared to those with exponential growth rates. Additionally, the inference time remains less than 80 ms for up to 36 BSs, making it practical for most scenarios.

### D. Adaptation of Proposed Framework to Disaggregated Architectures (Open RAN)

The energy-saving solution, FAMAC, developed in this paper, is well-suited for integration into novel disaggregated architectures such as Open Radio Access Network (O-RAN). In the O-RAN architecture, the control plane functions are centralized in a RAN Intelligent Controller (RIC), which consists of a Non-Real-Time (Non-RT) RIC and a Near-Real-Time (Near-RT) RIC [44]. FAMAC can be seamlessly integrated into the RIC as an rApp (RAN Application), enabling centralized decision-making for cell switching based on the global network view. The RIC collects traffic data from the distributed Radio Units (RUs) through standardized interfaces, such as the O1 interface connecting the Non-RT RIC to the Service Management and Orchestration (SMO) framework and the E2 interface connecting the Near-RT RIC to the E2 nodes.

The collected data is utilized by the FAMAC rApp to perform traffic prediction using a federated learning model, ensuring data privacy and reducing communication overhead. The predicted traffic patterns are then fed into a modified actor-critic deep RL algorithm, which makes intelligent decisions on cell switching to optimize energy consumption while maintaining the desired QoS. The optimal cell switching strategies determined by the FAMAC rApp are communicated back to the E2 nodes through the E2 interface, which then configure the RUs to switch off or on accordingly. This process is facilitated by the Open Fronthaul (O-FH) interface, which defines the control and data plane protocols between the RUs and the Distributed Units (DUs) in the O-RAN architecture. By leveraging the standardized interfaces and the disaggregated nature of the O-RAN architecture, FAMAC can be seamlessly integrated as an rApp within the RIC, allowing it to access the necessary network data, make intelligent decisions, and orchestrate the switching of RUs to achieve significant energy savings.

### E. Limitation and Potential Research Directions

Although this work proposes a novel and advanced cell switching methodology, there are limitations due to its scope. First, the impacts of frequent status changes of BSs on the total power consumption have not been considered, as this would require a different and more extensive problem formulation beyond the current scope. Second, users originally associated with switched-off BSs need to be offloaded to MBSs, resulting in handovers that may degrade their service quality. Since this work focuses on network-side power consumption, user-side handover and service quality issues have not been addressed. Lastly, this work did not implement user-specific propagation due to the use of real data (see Section V-D), thus the impacts of cell switching on received signal quality have not been examined. Therefore, further research is needed to address the aforementioned issues. Specifically, comprehensive modeling is required that considers both the impacts of handovers and power consumption due to the activation/deactivation of BSs. Additionally, investigating user-specific received signal strength analysis could help explore the energy-data rate trade-off.

## VII. CONCLUSION

In this work, we considered the energy optimization problem in 6G cellular networks comprising ultra-dense heterogeneous BS deployment to obtain a sustainable and cost-efficient network. To achieve this, we proposed a proactive and intelligent energy-saving framework in which the future traffic of the BSs is determined beforehand in a secure manner using FL. This is followed by the application of a modified version of the actor-critic DRL algorithm to determine the best cell switching policy that would lead to maximum energy savings in the network. The performance evaluations revealed that the proposed framework could save significant energy compared to state-of-the-art solutions while respecting the QoS constraints. In addition, it is scalable, which means that it can be applied to 6G networks even when their dimensions become large. In the future, we plan to consider various techniques for enhancing the offloading capacity of macro BS to give room for more SBSs to be switched off, thereby improving the amount of energy savings that can be obtained in the network.